\documentclass[]{pasj01}
\usepackage[switch,mathlines]{lineno}

\Received{04 June 2024}
\Accepted{17 July 2024}

\jyear{2024}

\newcommand{\Ni}{\ensuremath{^{56}\mathrm{Ni}}}

\newcommand{\Msun}{\ensuremath{\mathrm{M}_\odot}}

\newcommand{\Msunpyr}{\ensuremath{\Msun~\mathrm{yr^{-1}}}}
\newcommand{\kmps}{\ensuremath{\mathrm{km~s^{-1}}}}

\graphicspath{{./}{figures/}}

\DeclareAbbreviation\jcap{Journal of Cosmology and Astroparticle Phys.}

\usepackage{url}
 
\begin{document} 

\title{
Progenitor and explosion properties of SN~2023ixf estimated based on a light-curve model grid of Type~II supernovae
}

\author{Takashi J. \textsc{Moriya}\altaffilmark{1,2,3}}
\altaffiltext{1}{National Astronomical Observatory of Japan, National Institutes of Natural Sciences, 2-21-1 Osawa, Mitaka, Tokyo 181-8588, Japan}
\altaffiltext{2}{Graduate Institute for Advanced Studies, SOKENDAI, 2-21-1 Osawa, Mitaka, Tokyo 181-8588, Japan}
\altaffiltext{3}{School of Physics and Astronomy, Monash University, Clayton, VIC 3800, Australia}
\email{takashi.moriya@nao.ac.jp}

\author{Avinash \textsc{Singh}\altaffilmark{4,5}}
\altaffiltext{4}{Hiroshima Astrophysical Science Center, Hiroshima University, Higashi-Hiroshima, Hiroshima 739-8526, Japan}
\altaffiltext{5}{Department of Astronomy, The Oskar Klein Center, Stockholm University, AlbaNova University Center, SE 106 91 Stockholm, Sweden}
\email{avinash21292@gmail.com}



\maketitle

\begin{abstract}
We estimate the progenitor and explosion properties of the nearby Type~II SN~2023ixf using a synthetic model grid of Type~II supernova light curves. By comparing the light curves of SN~2023ixf with the pre-existing grid of Type~II supernovae containing about 228,000 models with different combinations of the progenitor and explosion properties, we obtain the $\chi^2$ value for every model and evaluate the properties of the models providing small values of $\chi^2$. We found that the light-curve models with the progenitor zero-age main-sequence mass of 10~\Msun, the explosion energy of $(2-3)\times 10^{51}~\mathrm{erg}$, the $^{56}\mathrm{Ni}$ mass of $0.04-0.06~\Msun$, the mass-loss rate of $10^{-3}-10^{-2}~\Msunpyr$ with a wind velocity of 10~\kmps, and the dense, confined circumstellar matter radius of $(6-10)\times 10^{14}~\mathrm{cm}$ match well to the observed light curves of SN~2023ixf. The photospheric velocity evolution of these models is also consistent with the observed velocity evolution. We note that the progenitor mass estimate could be affected by the adopted progenitor models. Although our parameter estimation is based on a pre-existing model grid and we do not perform any additional computations, the estimated parameters are consistent with those obtained by the detailed modeling of SN~2023ixf previously reported. This result shows that comparing the pre-existing model grid is a reasonable way to obtain a rough estimate for the properties of Type~II supernovae. This simple way to estimate the properties of Type~II supernovae will be essential in the Vera C. Rubin Observatory's Legacy Survey of Space and Time (LSST) era when thousands of Type~II supernovae are expected to be discovered yearly.
\end{abstract}


\section{Introduction}
Type~II supernovae (SNe) are explosions of massive stars that retain hydrogen-rich envelopes until the time of their explosion. Type~II SNe are the most frequently observed type of core-collapse SNe (e.g., \cite{perley2020}), and they provide vital information to understand the standard explosion mechanism of core-collapse SNe (e.g., \cite{burrows2021}). However, many mysteries related to Type~II SNe remain unsolved. For example, photometric and spectroscopic observations of Type~II SNe shortly after their explosion indicate the existence of confined circumstellar matter (CSM) around red supergiants (RSGs) that are about to explode (e.g., \cite{yaron2017,forster2018}). Concretely, optical light curves of Type~II SNe rise much faster than those predicted without considering the existence of a CSM \citep{gonzalez-gaitan2015}, and the fast rise can be explained by the existence of a dense, confined CSM around the progenitors of Type~II SNe (e.g., \cite{moriya2011,nagy2016,morozova2017}). In addition, SN~2020tlf showed a bright precursor lasting for 130~days, during which the dense, confined CSM could have likely formed \citep{jacobson-galan2022}. However, the physical mechanisms forming the dense, confined CSM have not been revealed.

SN~2023ixf is one of the closest Type~II SNe observed in the last decade, and their detailed observations provided us with valuable information on Type~II SNe and their progenitors. SN~2023ixf was discovered immediately after the explosion \citep{itagaki2023}, and its intensive follow-up observations were conducted in broad wavelengths (e.g., \cite{berger2023,grefenstette2023,sgro2023,hosseinzadeh2023,koenig2023,teja2023,jacobson-galan2023,hiramatsu2023,zhang2023,vasylyev2023,smith2023,bostroem2023,yamanaka2023,zimmerman2023,chandra2023,li2023,singh2024}). These observations revealed the existence of a dense, confined CSM in SN~2023ixf. SN~2023ixf also demonstrated that the future multi-messenger observations of nearby SNe will also be interesting (e.g., \cite{sarmah2023,guetta2023,kheirandish2023,murase2023,ravensburg2024}). The progenitor of SN~2023ixf has been identified, but its nature is actively discussed (e.g., \cite{kilpatrick2023,jencson2023,pledger2023,flinner2023,vandyk2023,niu2023,qin2023,dong2023,liu2023,neustadt2024,xiang2024}).

While nearby Type~II SNe like SN~2023ixf can only appear once in a decade or so, thousands of distant Type~II SNe are predicted to be discovered in a year in the era of the Vera C. Rubin Observatory's Legacy Survey of Space and Time (LSST, \cite{lsst2009}). It will be impossible to perform intensive follow-up observations of such a large number of Type~II SNe individually as conducted for SN~2023ixf. A systematic approach is required to infer the properties of a large number of Type~II SNe, for which only limited photometric information is expected to be available. To estimate the physical properties of a large number of Type~II SNe, \citet{moriya2023} constructed a model grid of Type~II SNe containing about 228,000 explosion models with different combinations of progenitors, explosion properties, and CSM properties. This model grid has been used to estimate the properties of a large number of Type~II SNe discovered by ongoing extensive transient surveys such as Zwickey Transient Facility (ZTF, e.g., \cite{subrayan2023,silva2024}).

In this paper, we present our estimates of the progenitor and explosion properties of SN~2023ixf based on the model grid of \citet{moriya2023}. We discuss the distribution of the parameters estimated by the model grid and compare our parameter estimates with those estimated by other studies. This study allows us to test the ability of the model grid to estimate the properties of Type~II SNe by using the well-observed Type~II SN~2023ixf. We estimate the properties of SN~2023ixf by using only the pre-computed model grid in \citet{moriya2023} without additional calculations. Using the pre-computed model grid, we can quickly identify the rough nature of SN~2023ixf, based on which more detailed modeling of its light curves and spectra can be performed.

The rest of this paper is organized as follows. We first present our methods to compare the model grid and the observations of SN~2023ixf in Section~\ref{sec:method}. We present the results of the model comparison and our parameter estimation in Section~\ref{sec:results}. We discuss our results in Section~\ref{sec:discussion} and conclude this paper in Section~\ref{sec:conclusions}.

\begin{figure*}
 \begin{center}
  \includegraphics[width=\columnwidth]{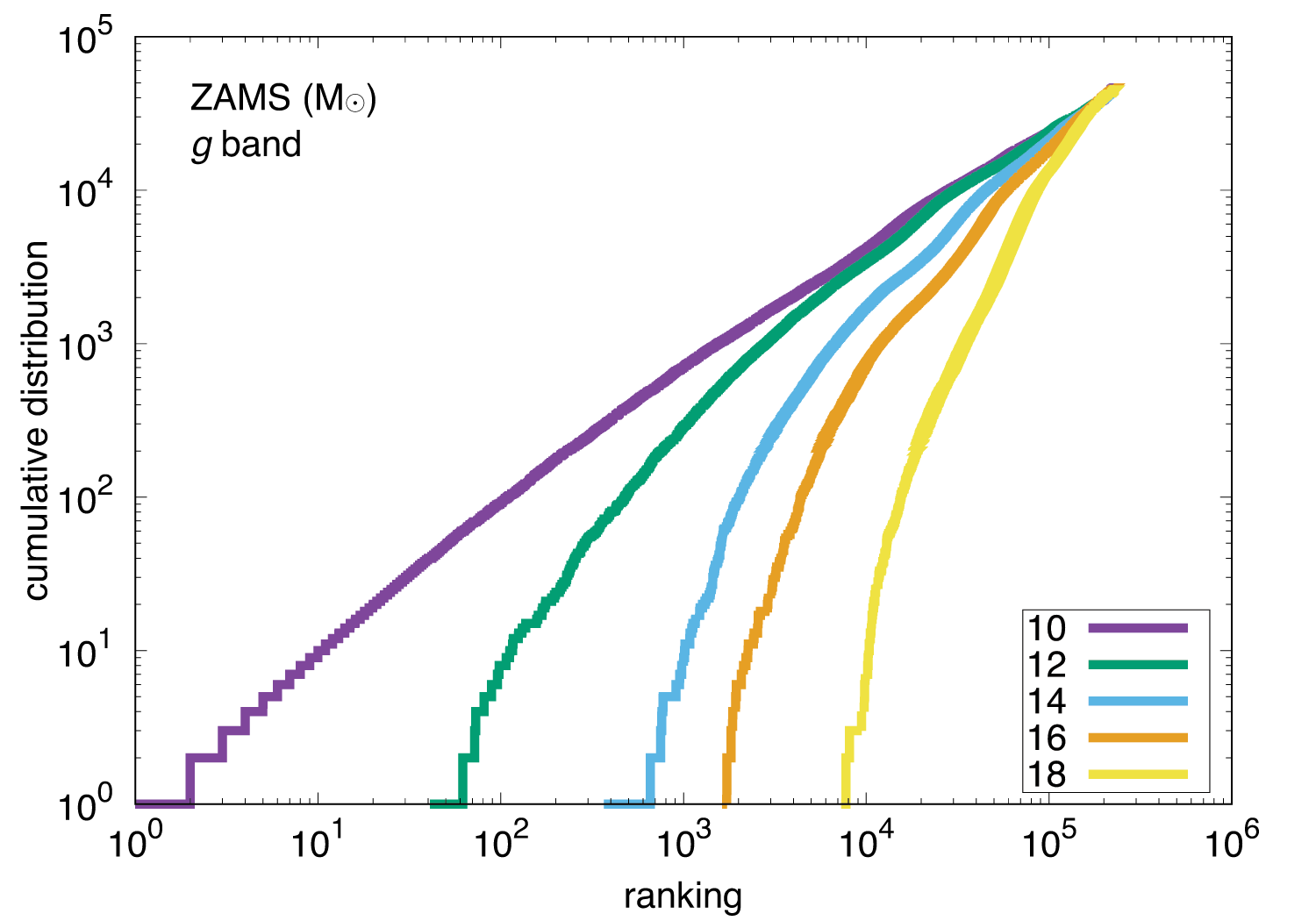}
  \includegraphics[width=\columnwidth]{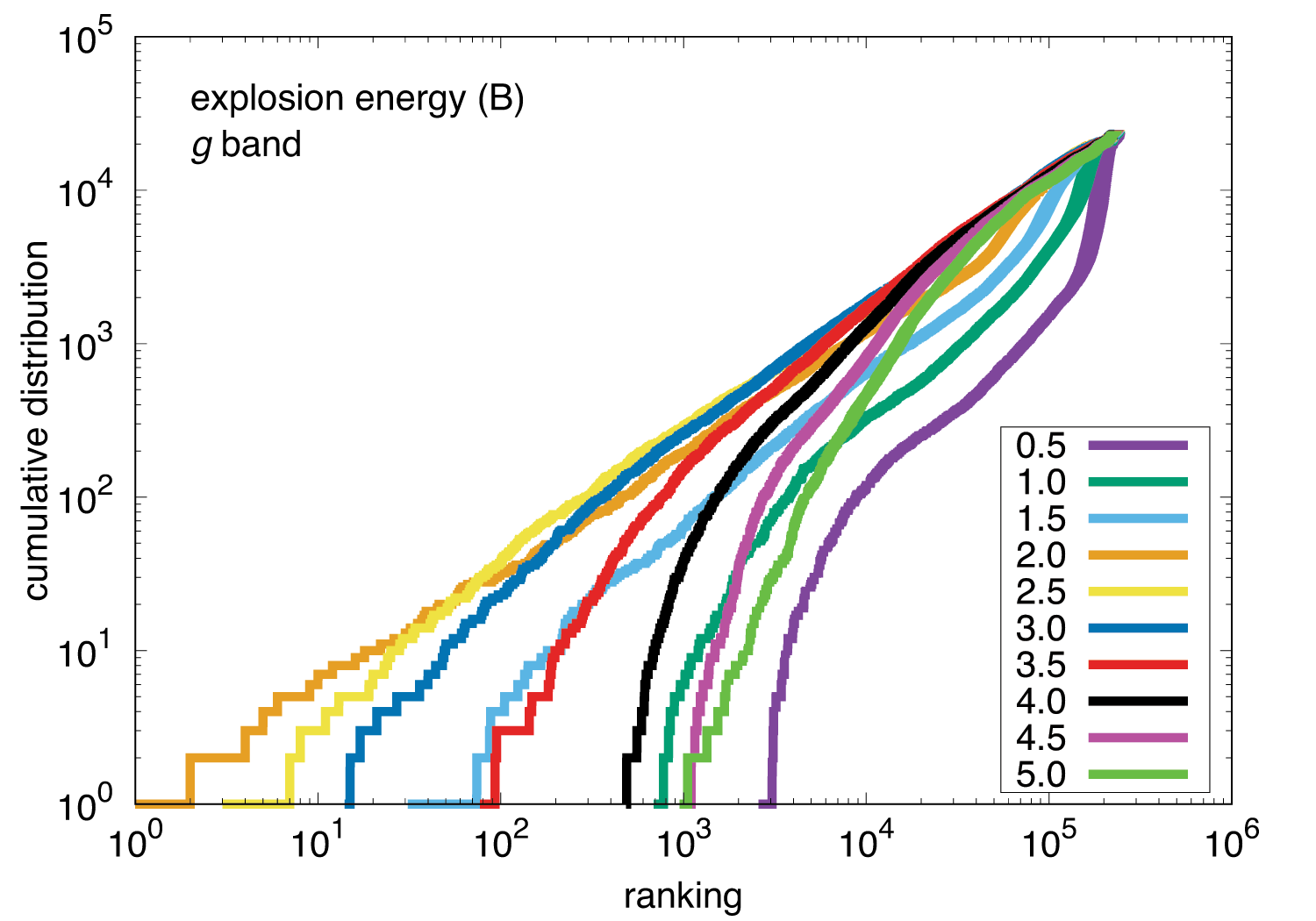}
  \includegraphics[width=\columnwidth]{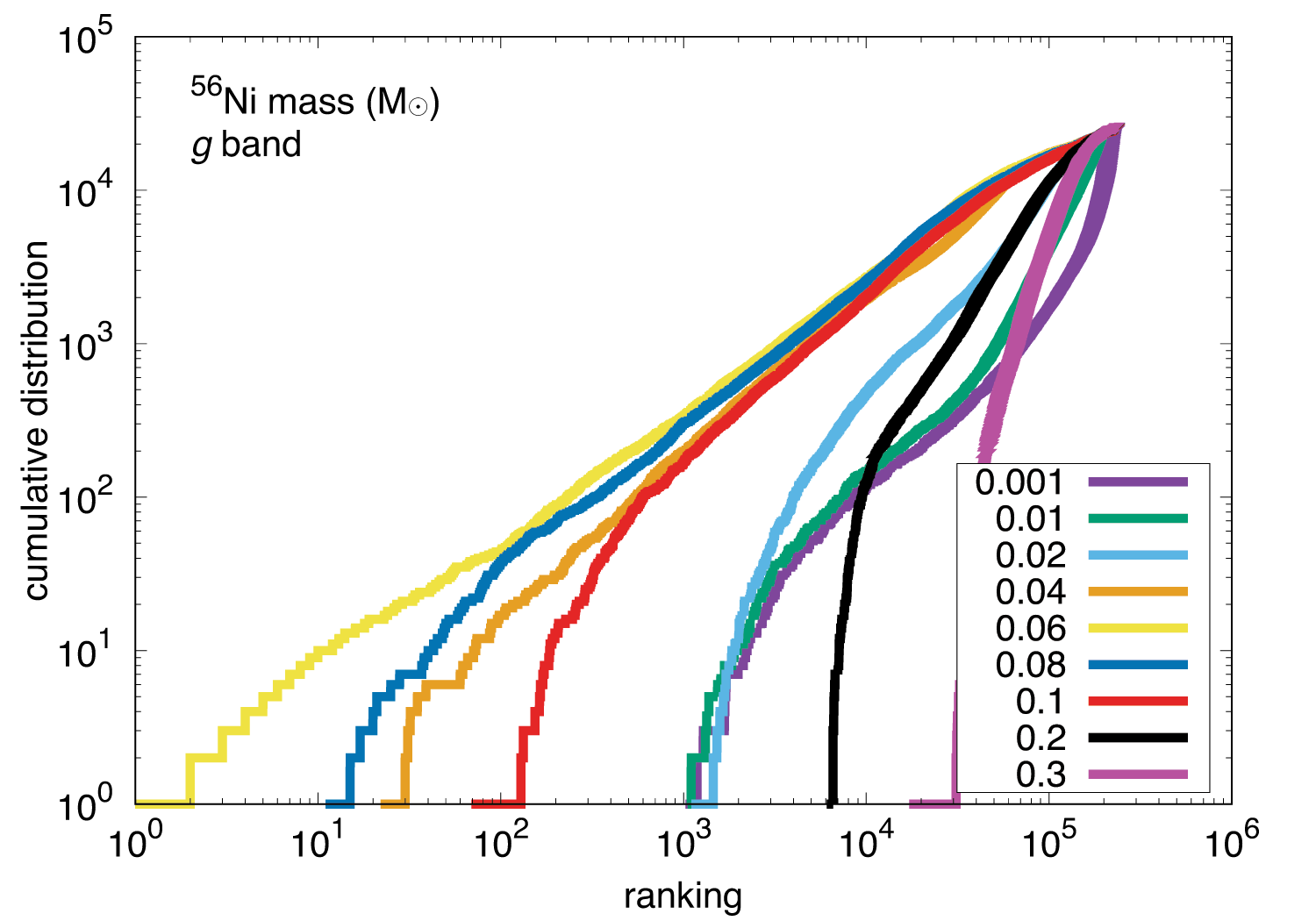}
  \includegraphics[width=\columnwidth]{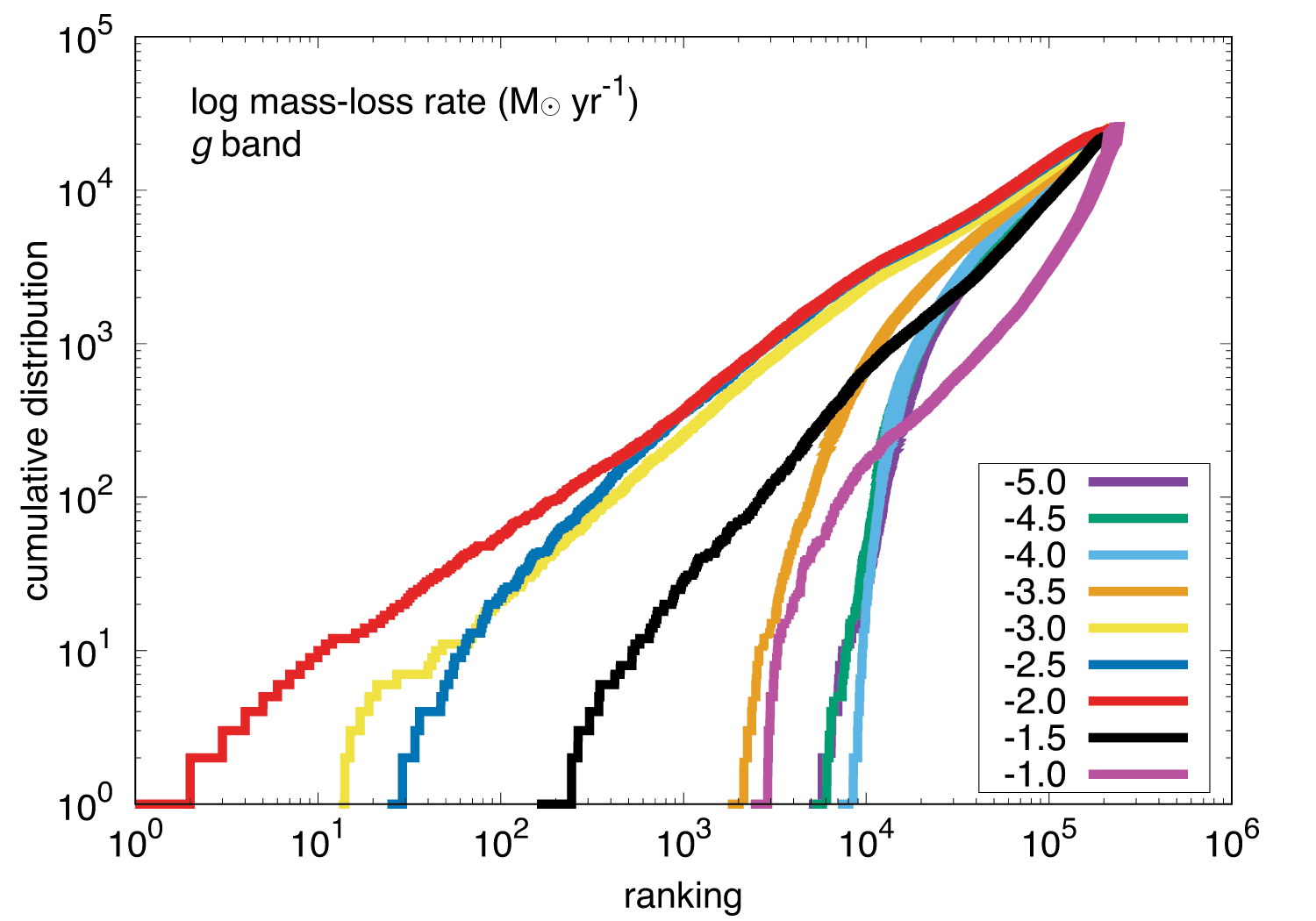}
  \includegraphics[width=\columnwidth]{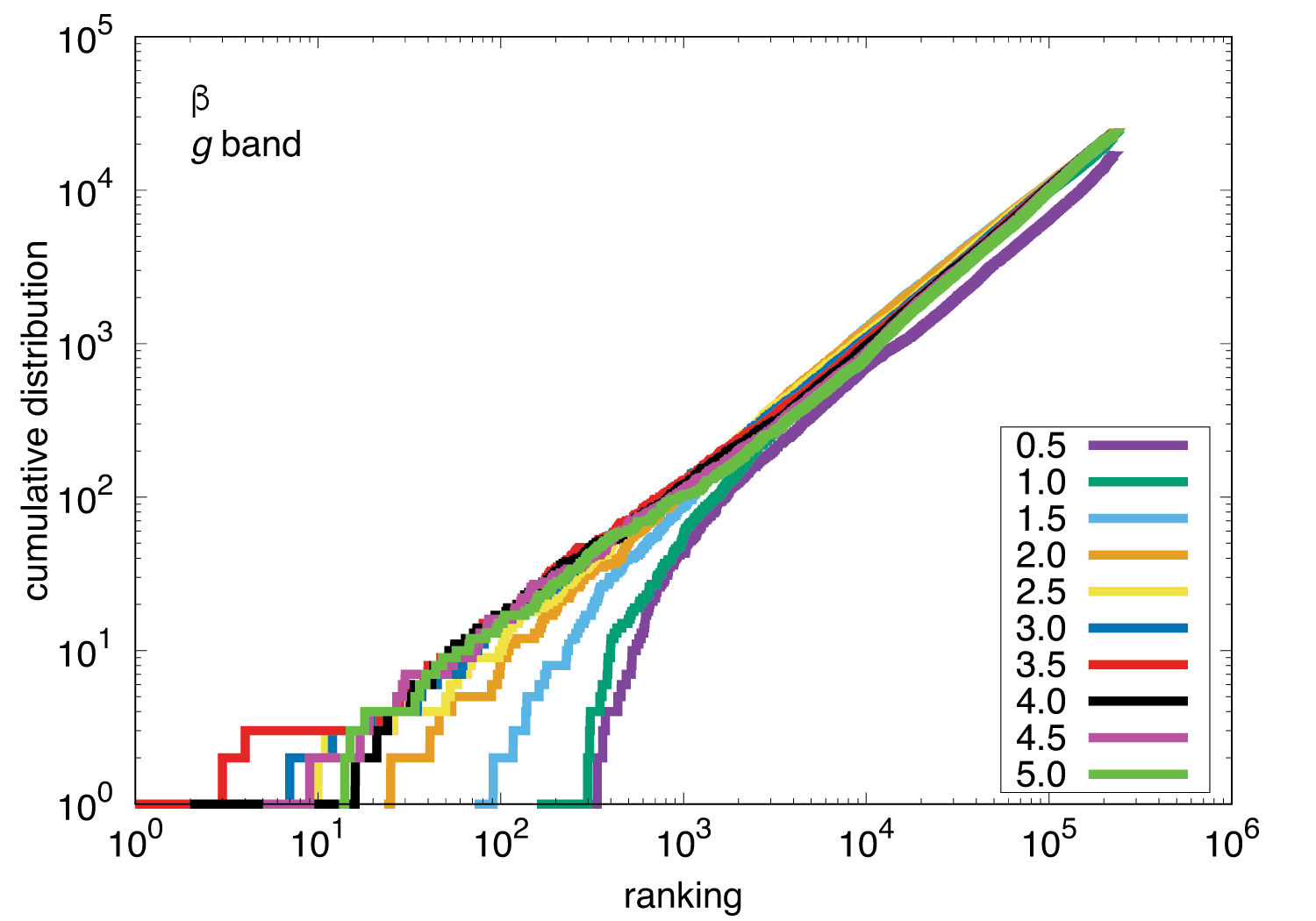}
  \includegraphics[width=\columnwidth]{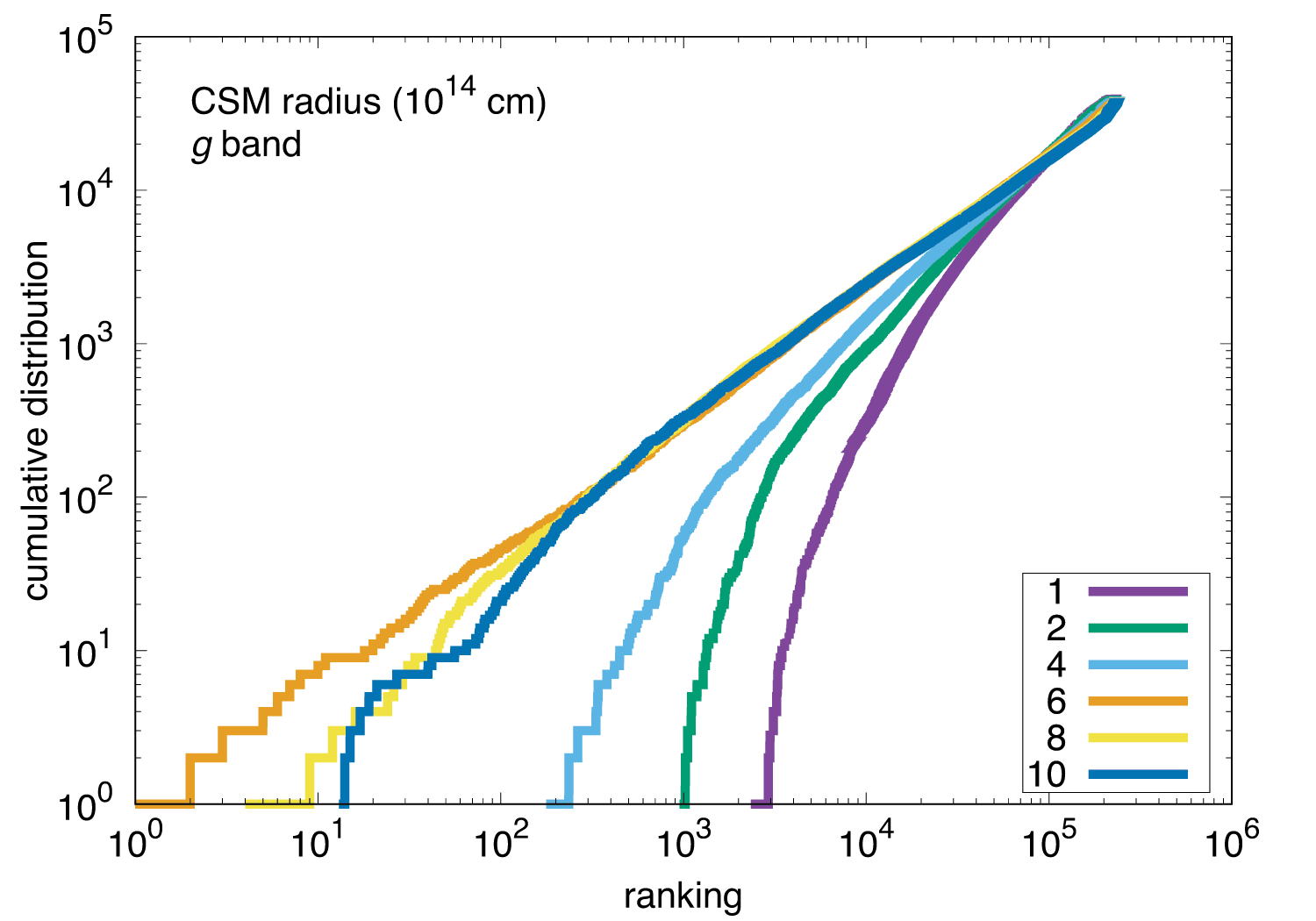}  
 \end{center}
\caption{
Cumulative ranking distributions of the properties of SN~2023ixf. Only the \textit{g}~band photometry is used to obtain the ranking distribution in this figure.
}\label{fig:g}
\end{figure*}

\begin{figure}
 \begin{center}
  \includegraphics[width=\columnwidth]{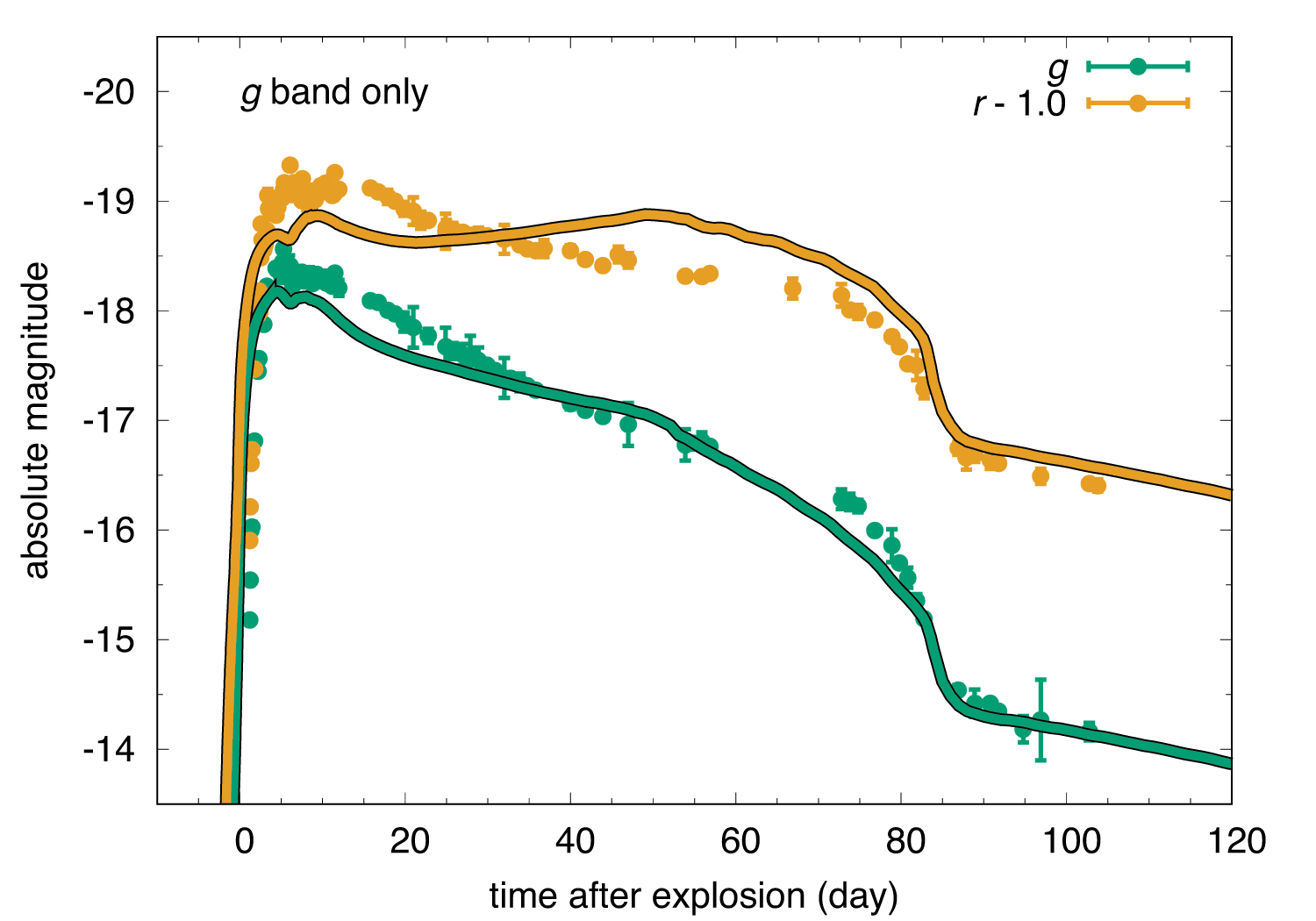}
  \includegraphics[width=\columnwidth]{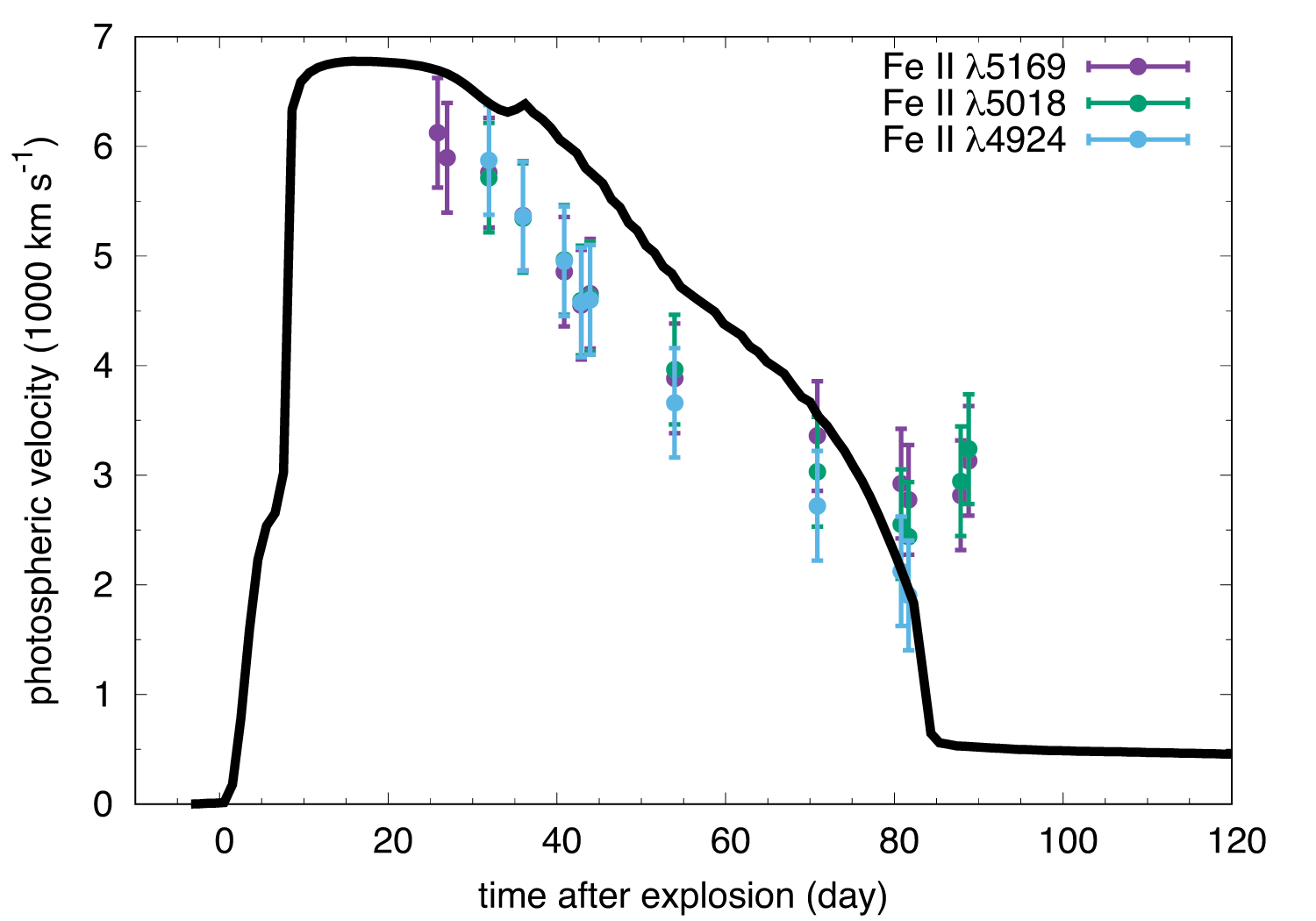}
 \end{center}
\caption{
The light-curve and photospheric velocity evolution of the model providing the least $\chi^2$ when we only use the \textit{g}~band photometry data to compare with the model grid.
}\label{fig:gbest}
\end{figure}

\begin{figure}[t]
 \begin{center}
  \includegraphics[width=\columnwidth]{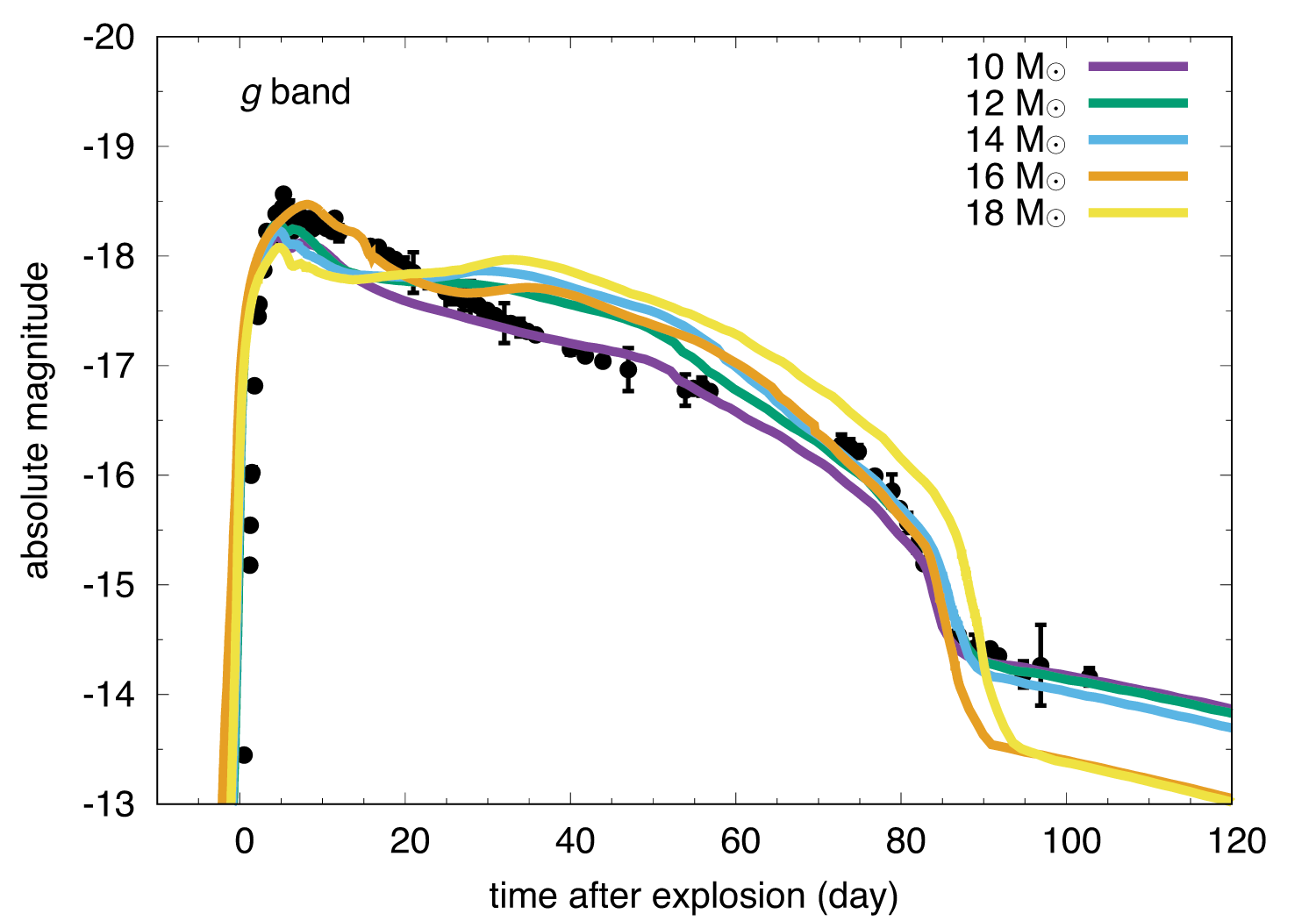}
 \end{center}
\caption{
The \textit{g} band light curves of the models providing the least $\chi^2$ for each progenitor model. \\
}\label{fig:gzams}
\end{figure}

\begin{figure*}
 \begin{center}
  \includegraphics[width=\columnwidth]{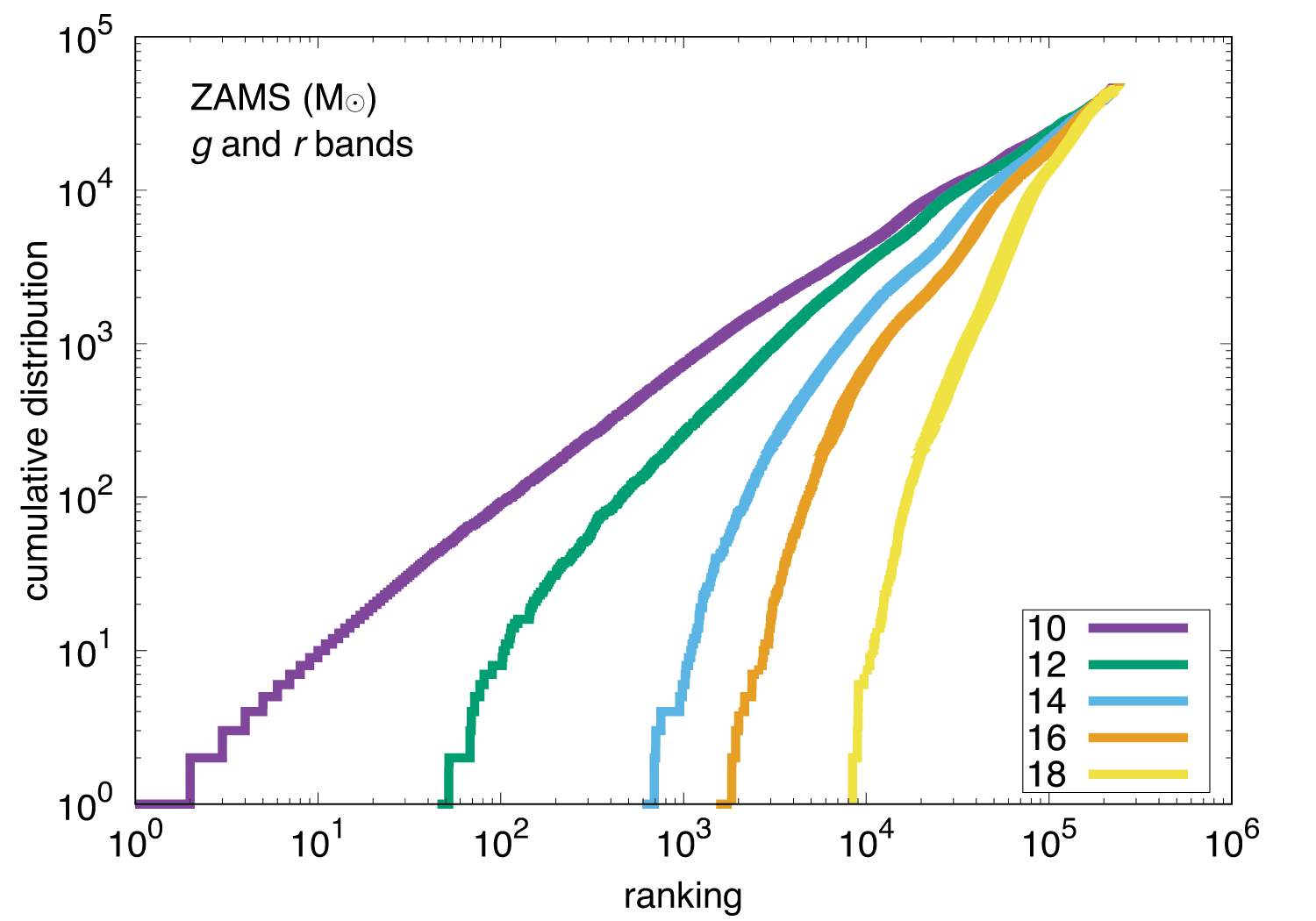}
  \includegraphics[width=\columnwidth]{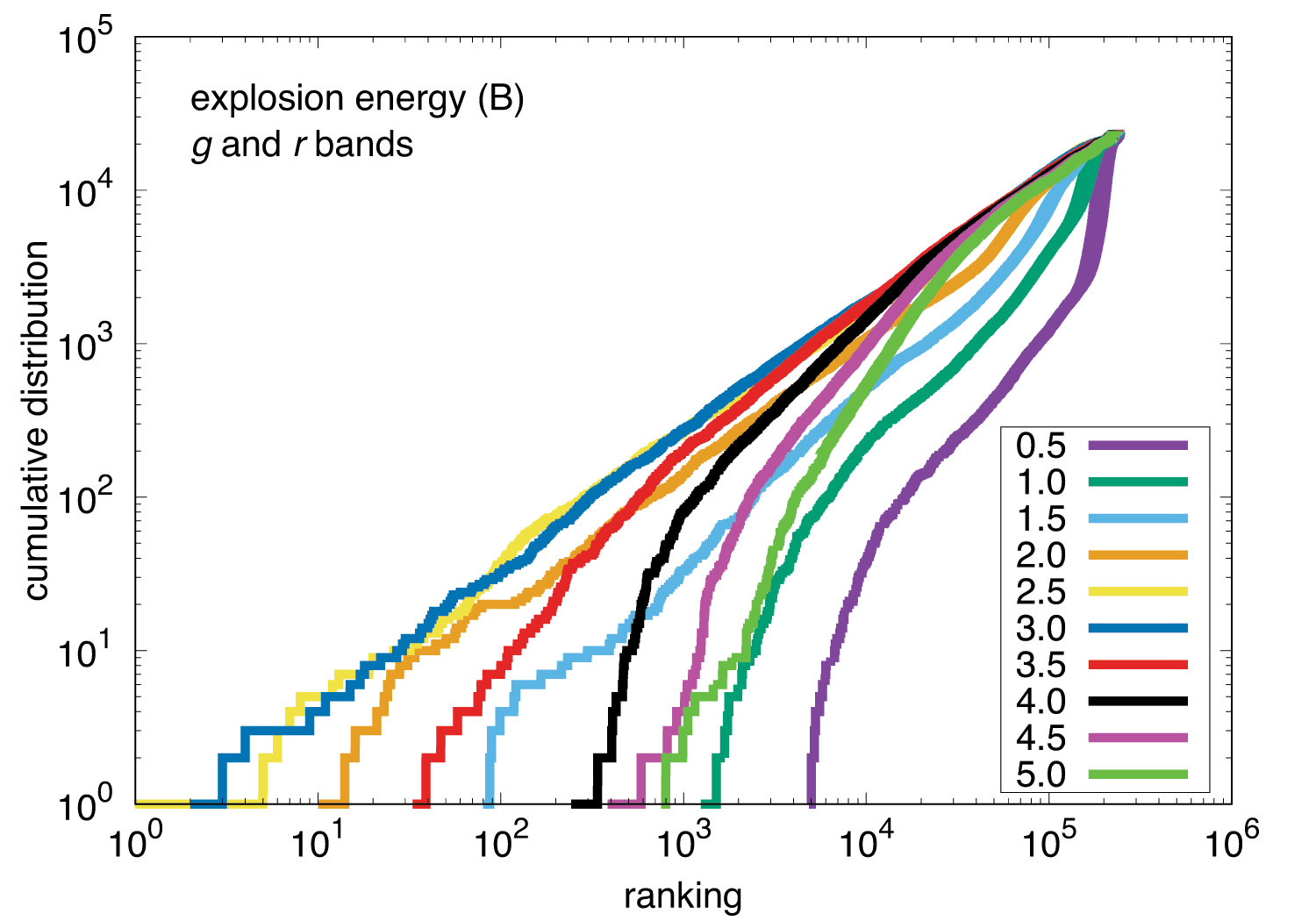}
  \includegraphics[width=\columnwidth]{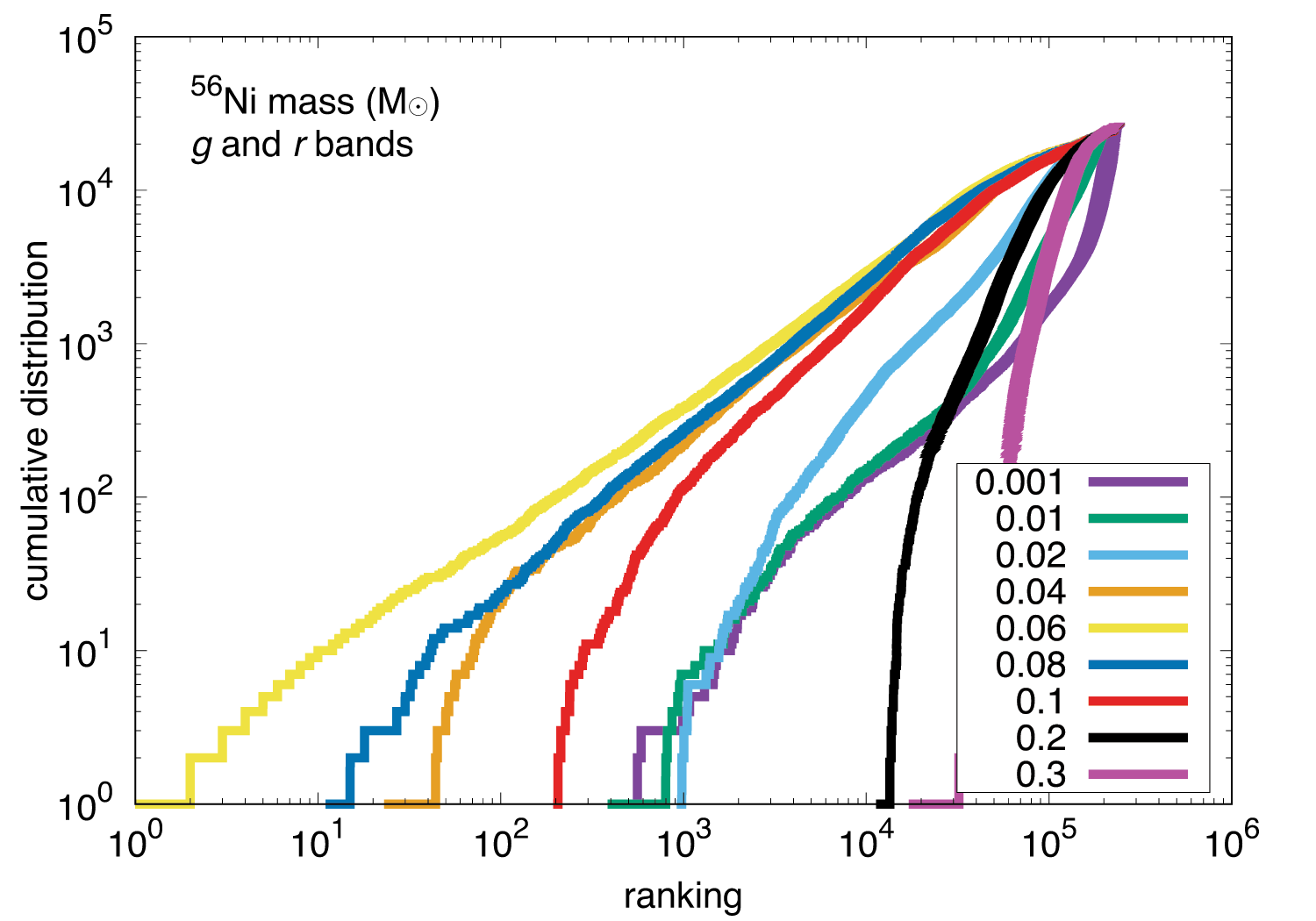}
  \includegraphics[width=\columnwidth]{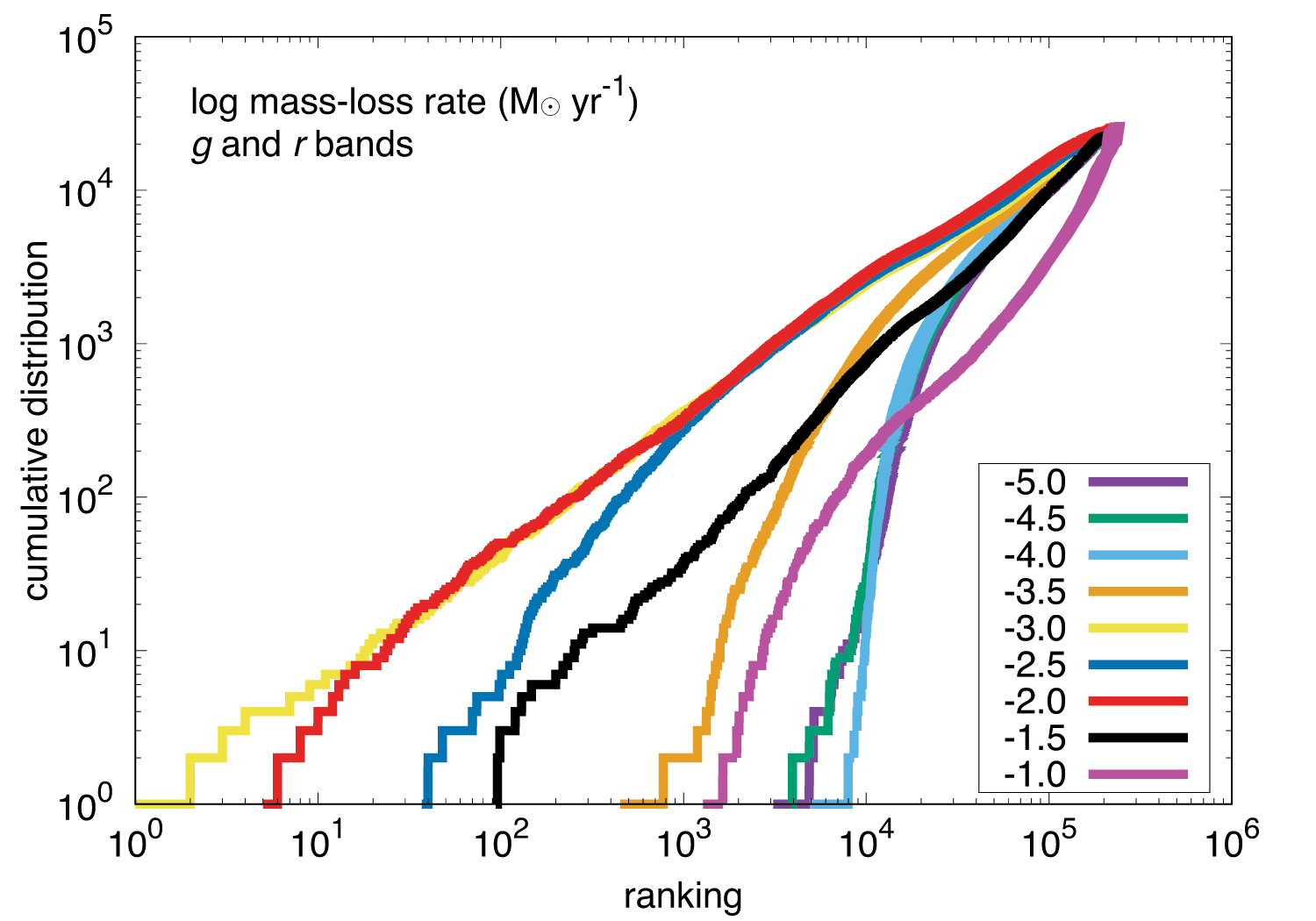}
  \includegraphics[width=\columnwidth]{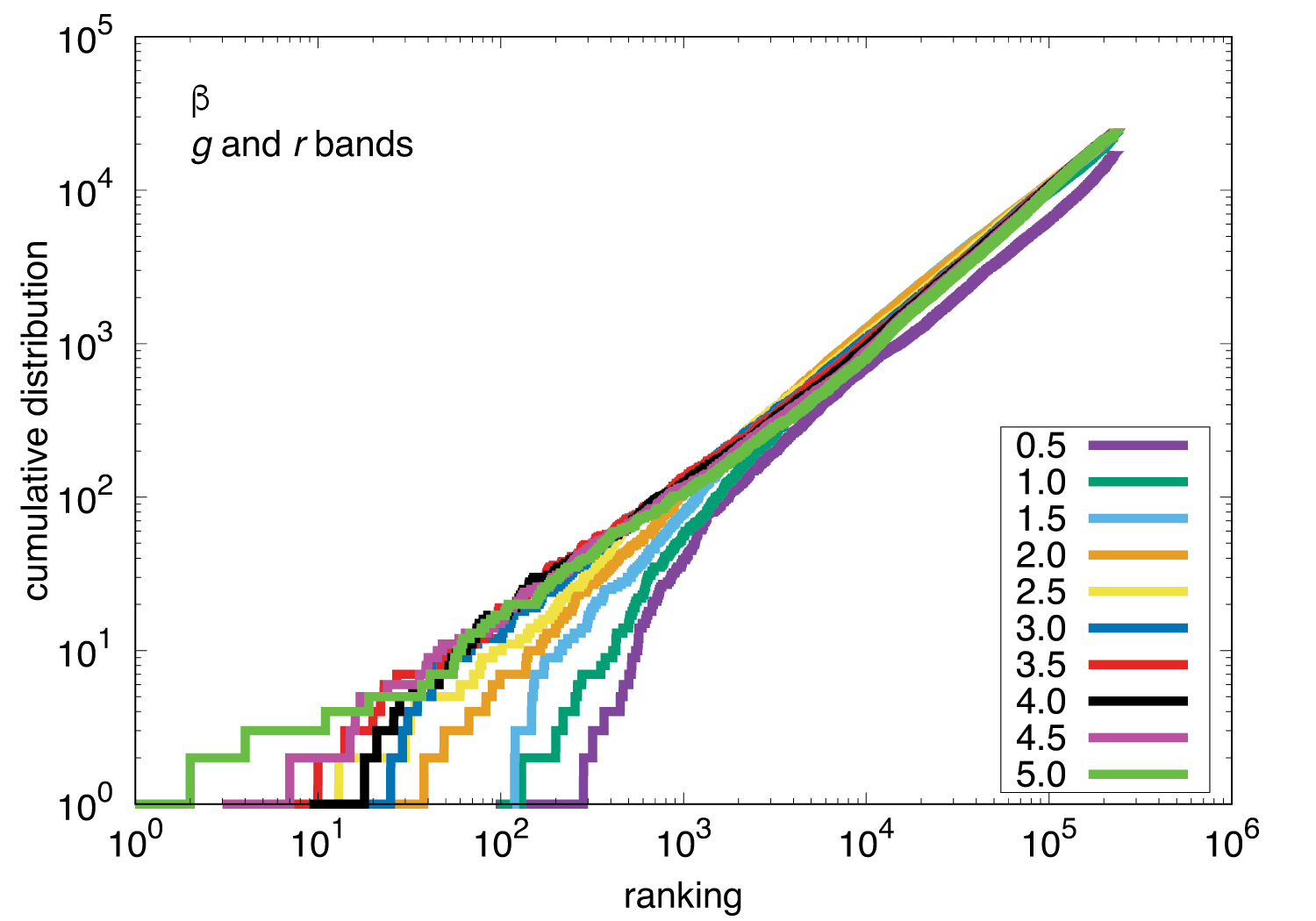}
  \includegraphics[width=\columnwidth]{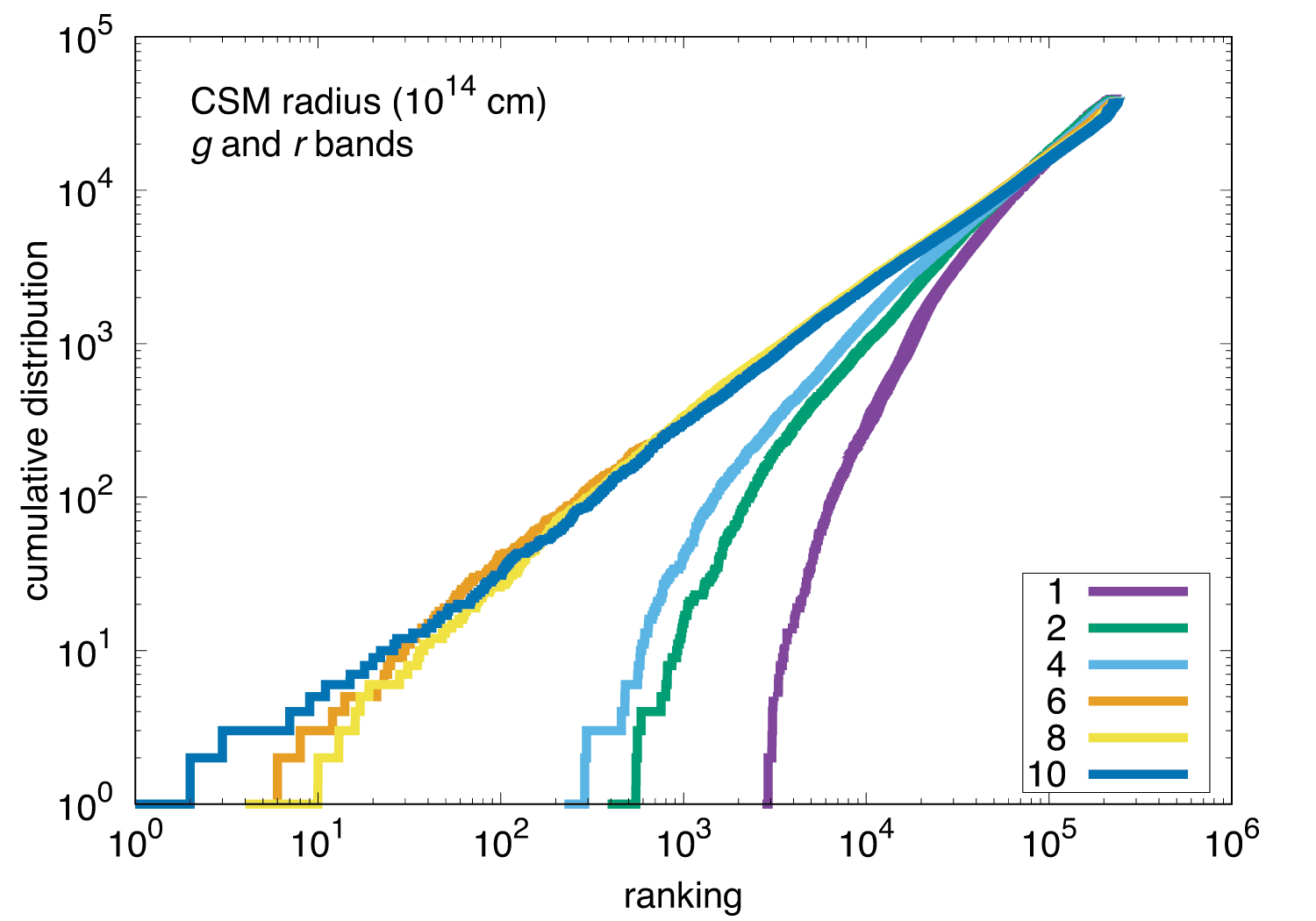}  
 \end{center}
\caption{
Cumulative ranking distributions of the properties of SN~2023ixf. The \textit{g} and \textit{r} band photometry are both used to obtain the ranking distribution in this figure.
}\label{fig:gr}
\end{figure*}

\begin{figure}
 \begin{center}
  \includegraphics[width=\columnwidth]{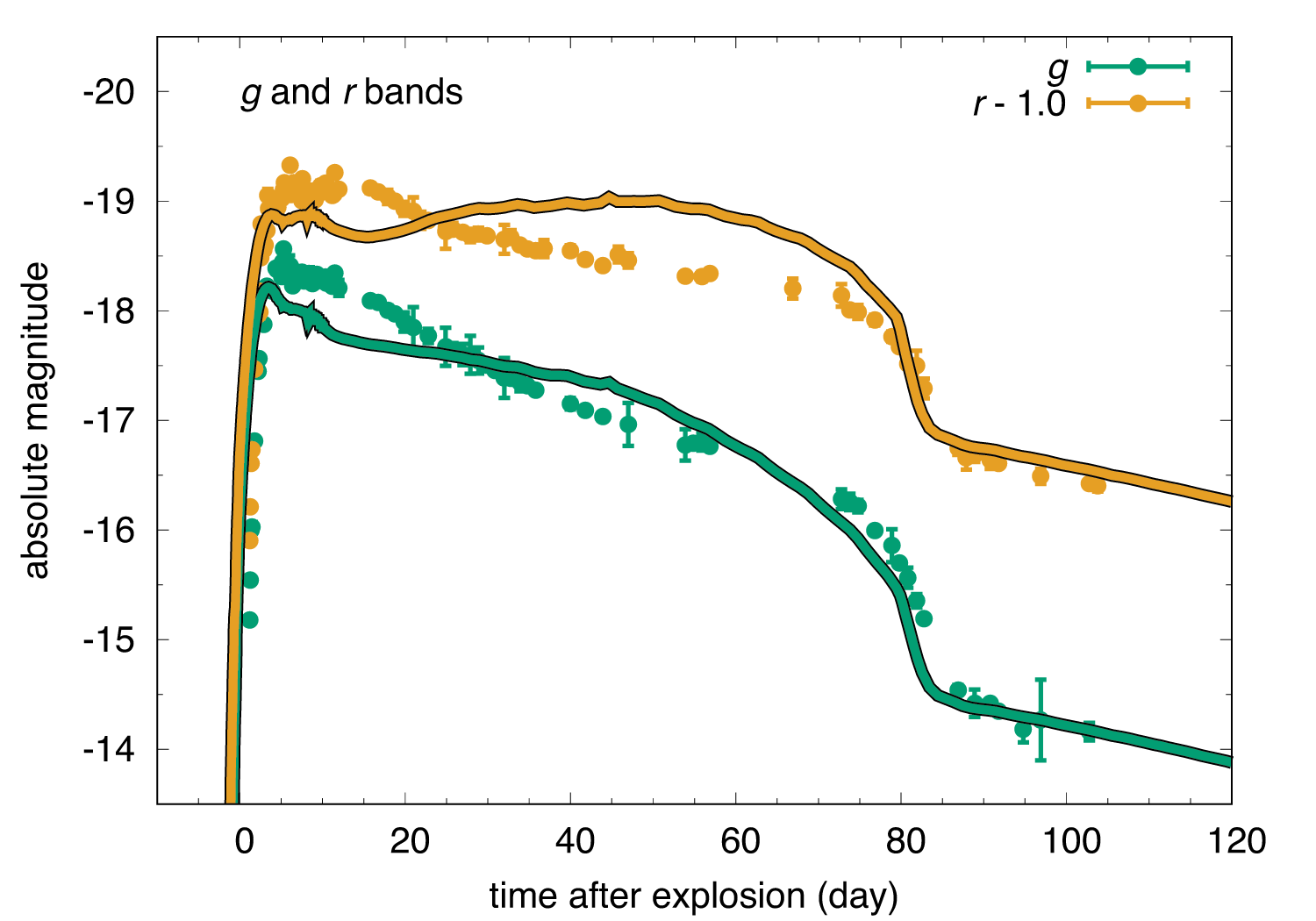}
  \includegraphics[width=\columnwidth]{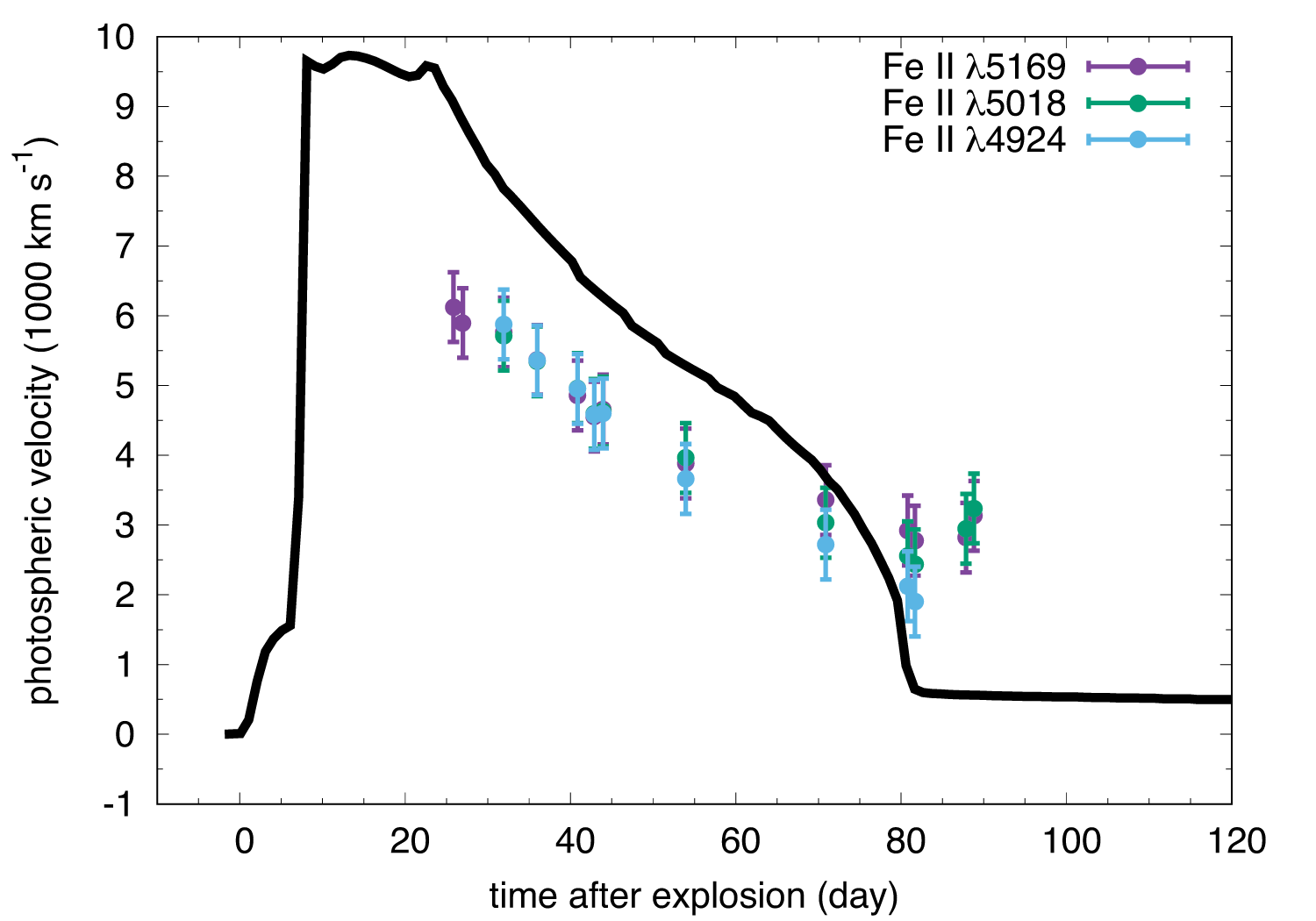}
 \end{center}
\caption{The light-curve and photospheric velocity evolution of the model that provides the least $\chi^2$ when we use the \textit{g} and \textit{r}~band photometry data to compare with the model grid.}\label{fig:grbest}
\end{figure}

\begin{figure*}
 \begin{center}
  \includegraphics[width=\columnwidth]{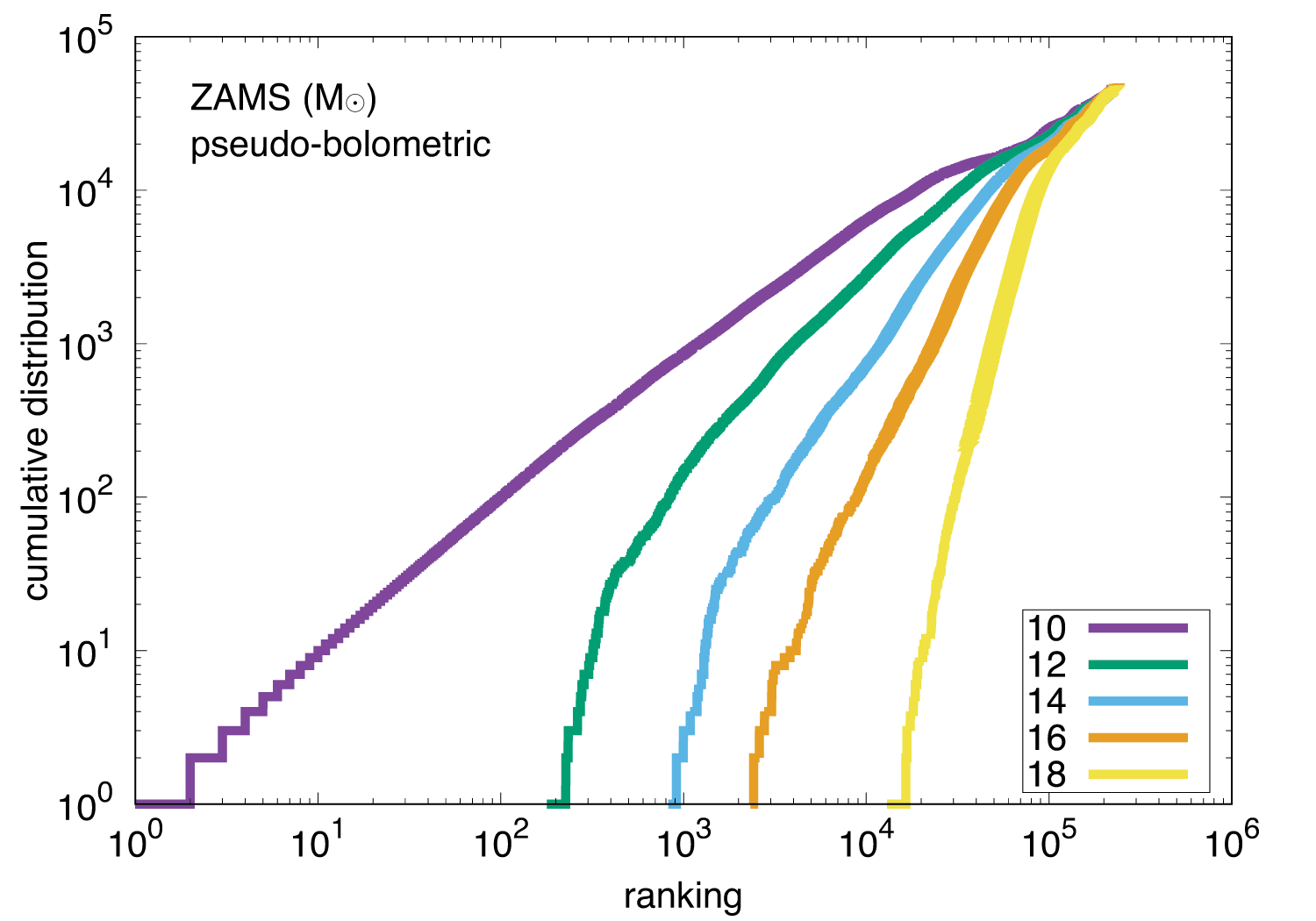}
  \includegraphics[width=\columnwidth]{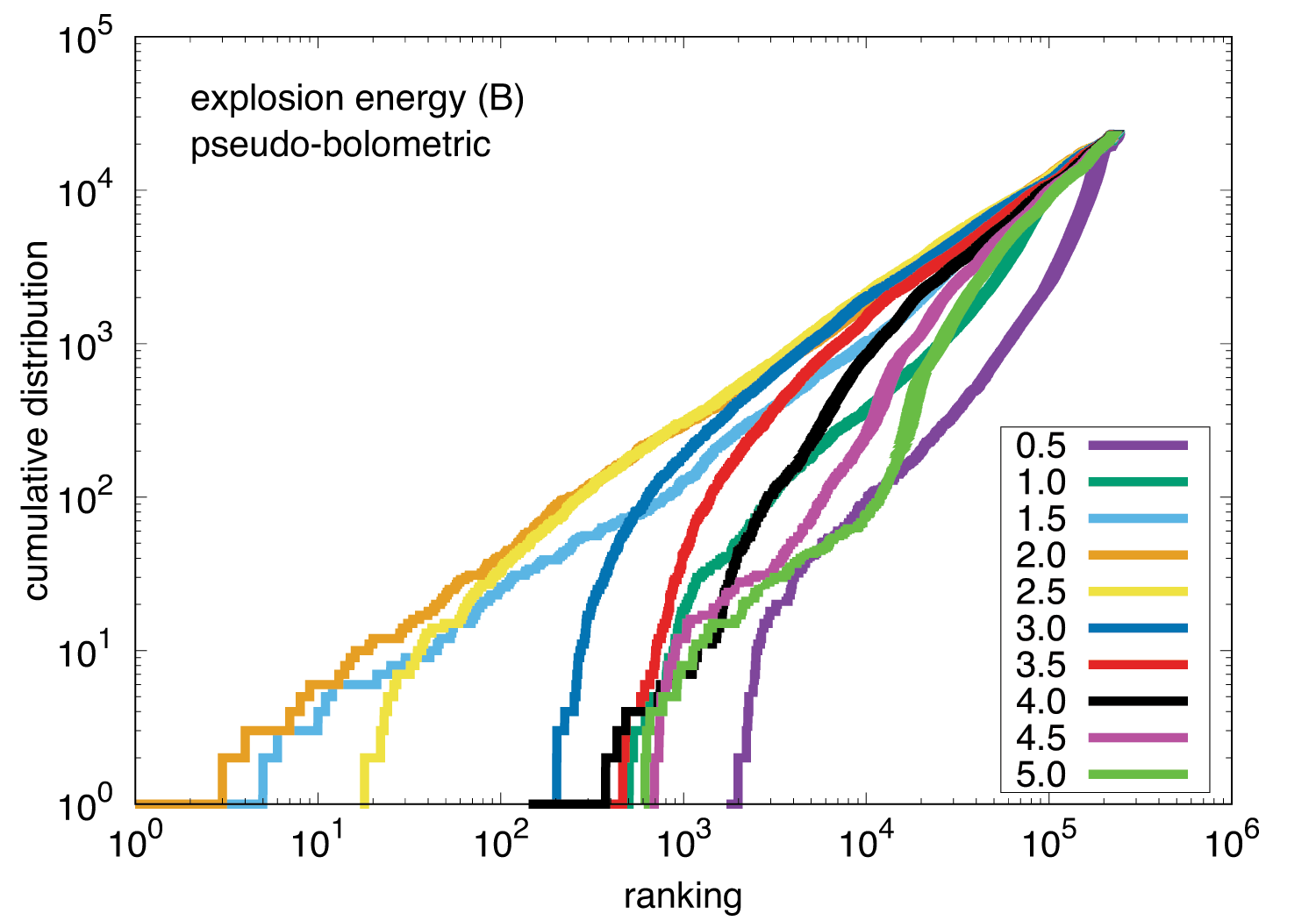}
  \includegraphics[width=\columnwidth]{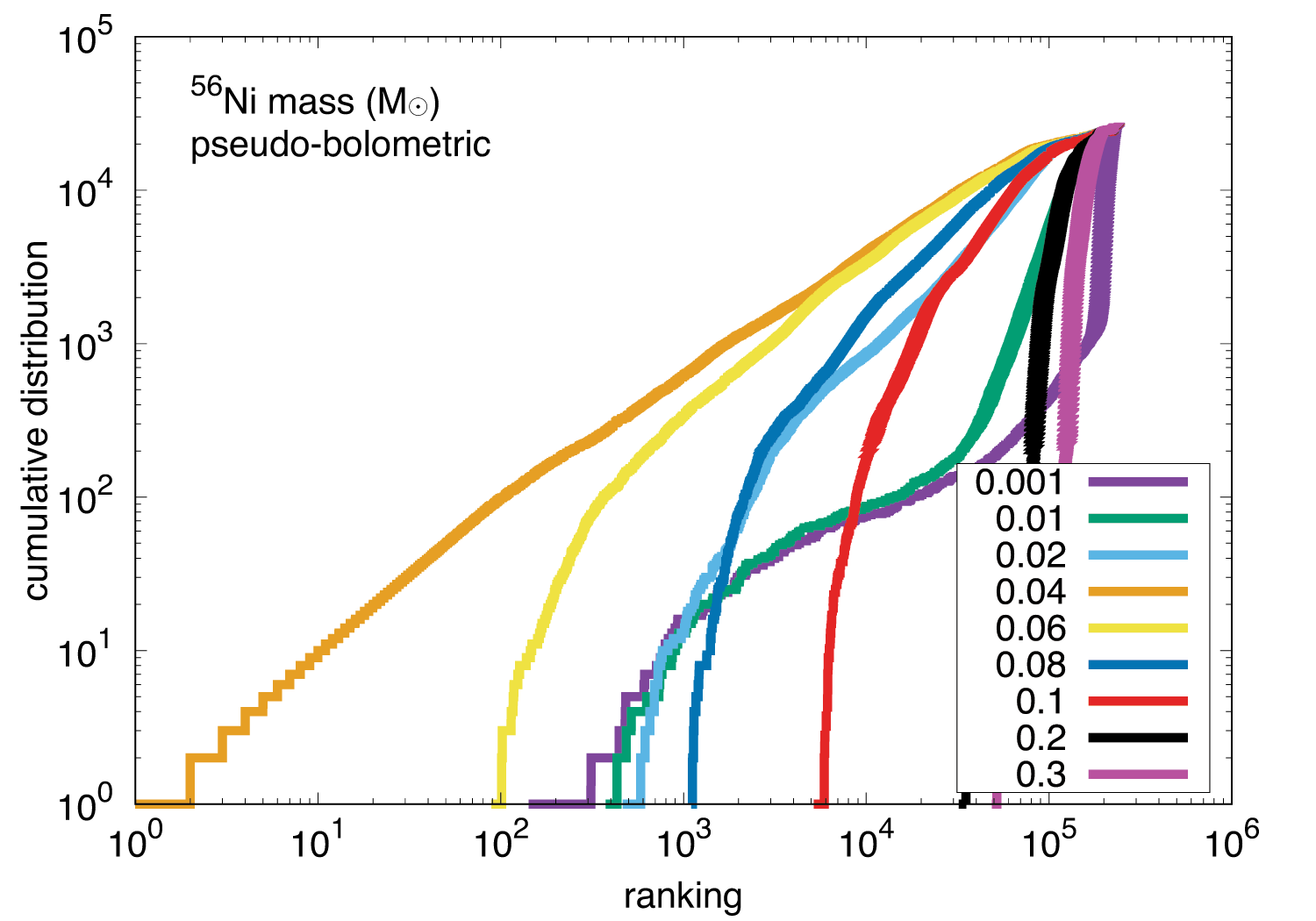}
  \includegraphics[width=\columnwidth]{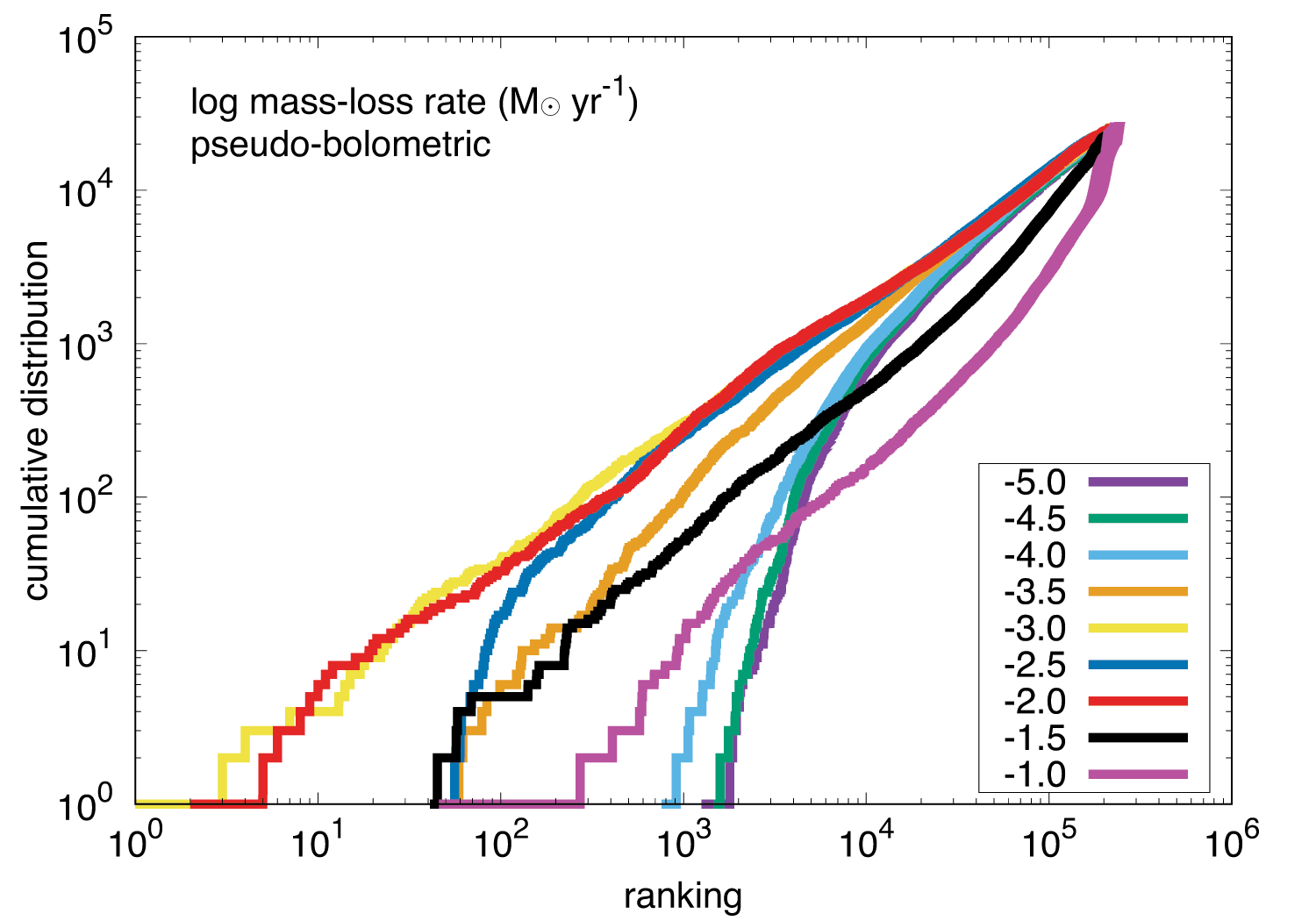}
  \includegraphics[width=\columnwidth]{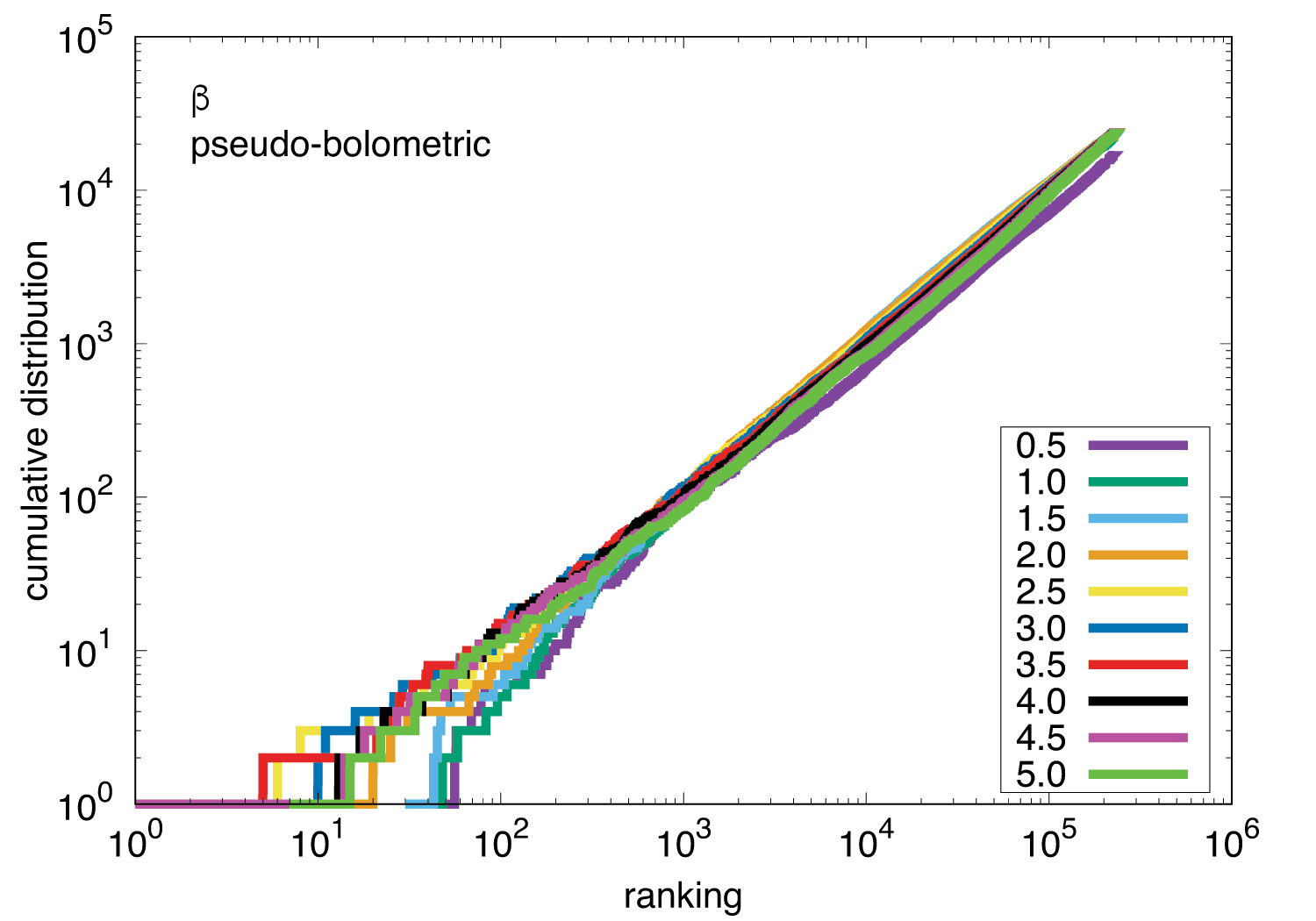}
  \includegraphics[width=\columnwidth]{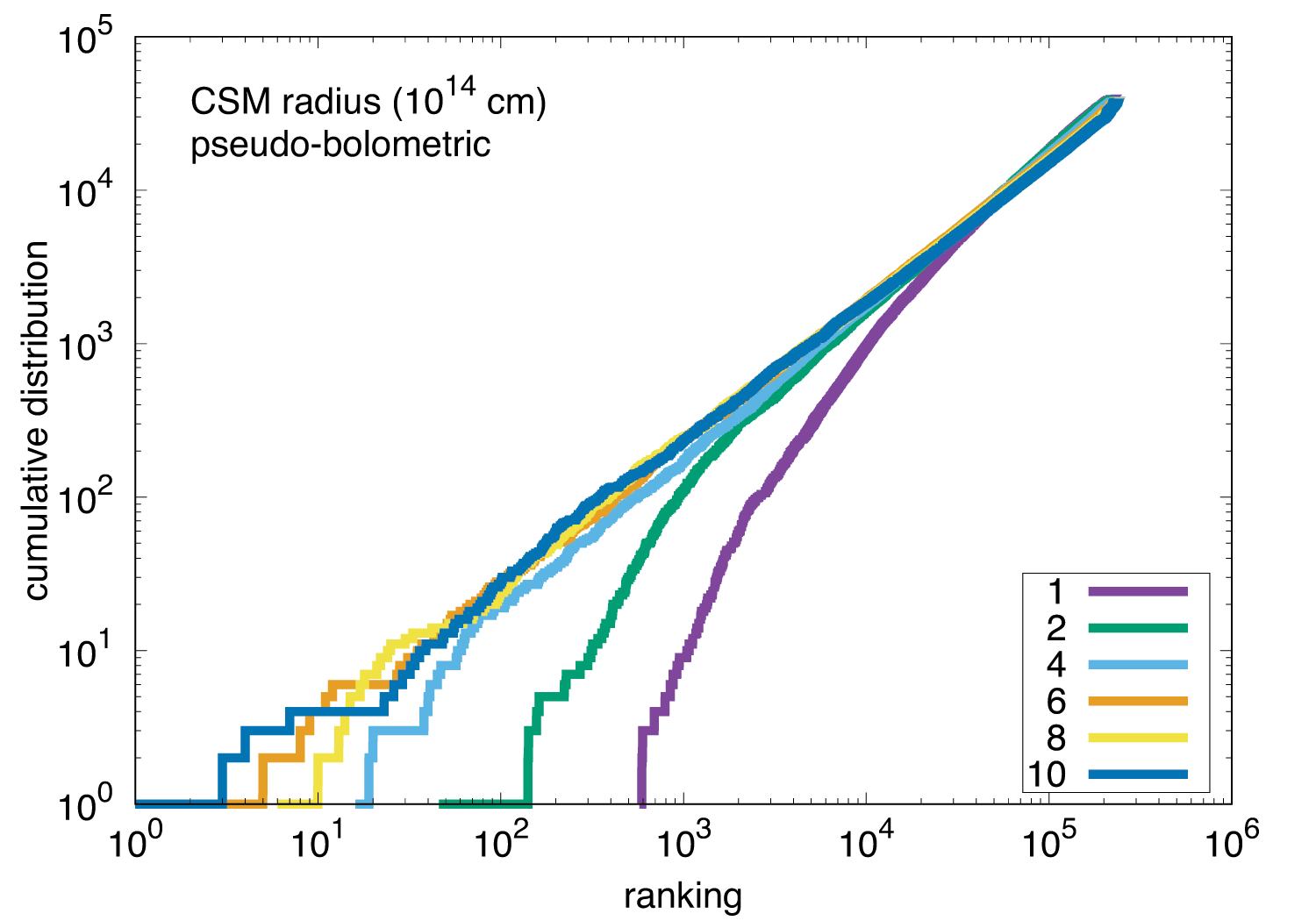}  
 \end{center}
\caption{
Cumulative ranking distributions of the properties of SN~2023ixf. The pseudo-bolometric light curve is used to obtain the ranking distribution in this figure.
}\label{fig:bol}
\end{figure*}

\begin{figure}
 \begin{center}
  \includegraphics[width=\columnwidth]{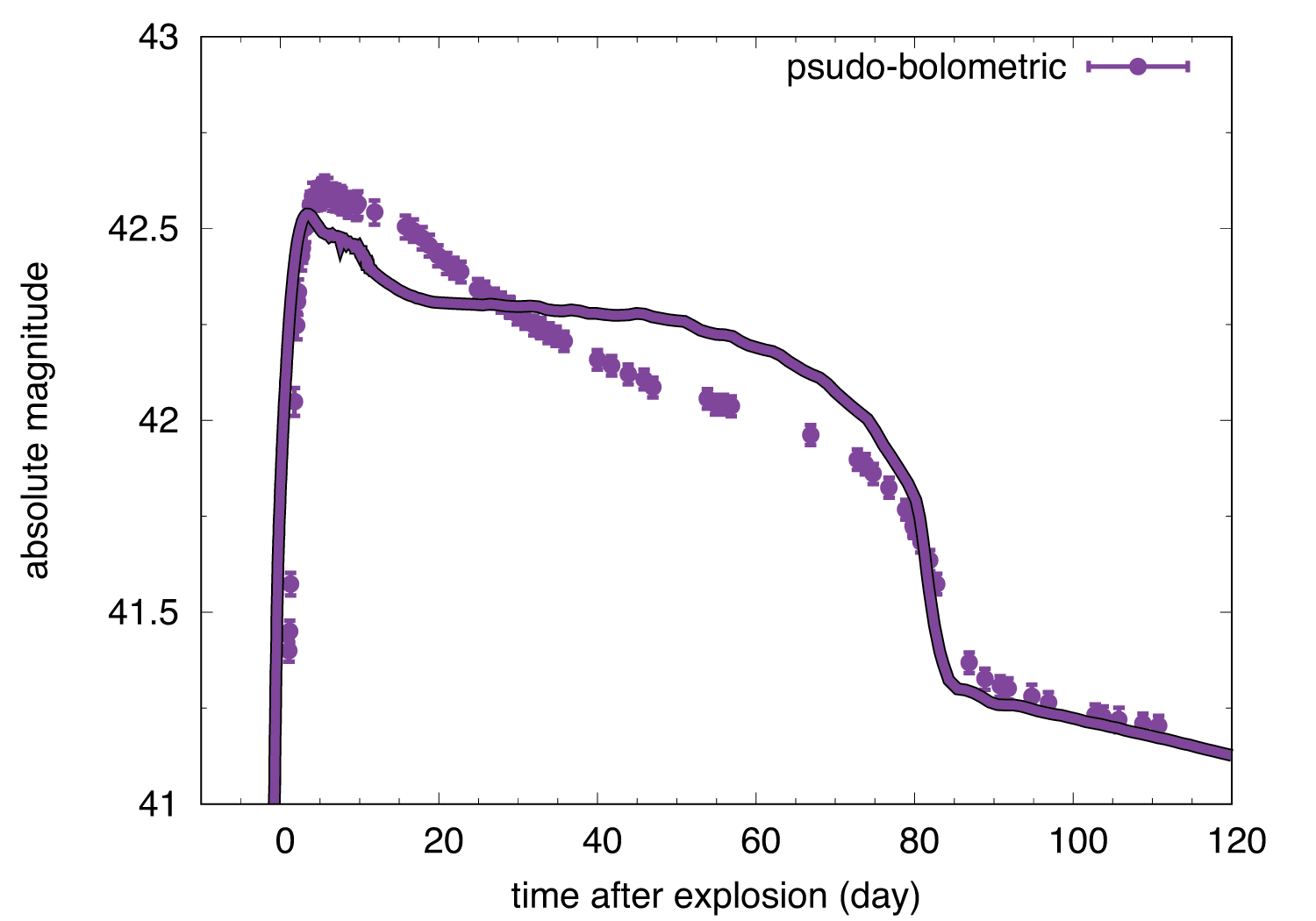}
  \includegraphics[width=\columnwidth]{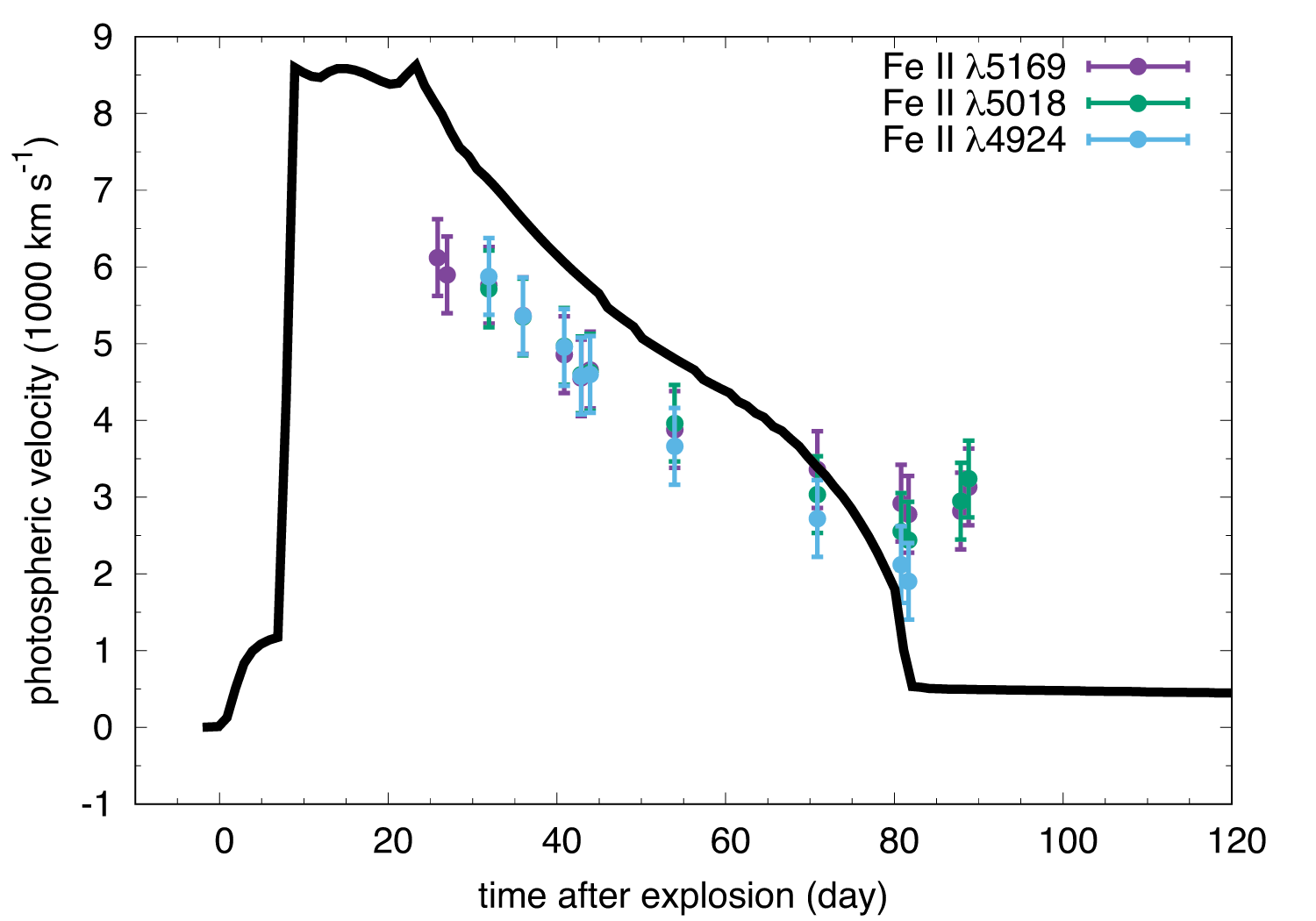}
 \end{center}
\caption{
The light-curve and photospheric velocity evolution of the model which provides the least $\chi^2$ when we compare the pseudo-bolometric light curve with the model grid.
}\label{fig:bolbest}
\end{figure}

\section{Methods}\label{sec:method}
\subsection{Model grid}
In this study, we briefly introduce the model grid of Type~II SNe used for the parameter estimation of SN~2023ixf. The full explanation of the model grid is available in \citet{moriya2023}, and the data of the model grid are available at \url{https://doi.org/10.5061/dryad.pnvx0k6sj}.

The model grid is constructed using the progenitor model from \citet{sukhbold2016}. The progenitor zero-age main-sequence (ZAMS) mass of the grid ranges from 10~\Msun\ to 18~\Msun. The dense CSM is attached artificially to the progenitor models. The mass-loss rates $\dot{M}$ in the model grid range from $10^{-5}~\Msunpyr$ to $10^{-1}~\Msunpyr$, and the CSM radii in the model grid range from $10^{14}~\mathrm{cm}$ to $10^{15}~\mathrm{cm}$. The CSM density $\rho_\mathrm{CSM}$ is assumed to follow
\begin{equation}
    \rho_\mathrm{CSM} = \frac{\dot{M}}{4\pi v_\mathrm{wind}(r)} r^{-2}, \label{eq:rho_csm}
\end{equation}
from the mass conservation. The wind velocity $v_\mathrm{wind}(r)$ is assumed to follow the $\beta$-law wind velocity profile (e.g., \cite{moriya2017,moriya2018})
\begin{equation}
    v_\mathrm{wind} (r) = v_0 + (v_\infty - v_0)\left(1-\frac{R_0}{r}\right)^\beta,
\end{equation}
where $v_0$ is the initial wind velocity at the stellar surface, $v_\infty$ is the terminal wind velocity, and $R_0$ is the progenitor radius. We set the terminal wind velocity to be $10~\kmps$. We adopt $\beta$ of $0.5-5$. $v_0$ is chosen to connect the progenitor to the CSM smoothly.

The light-curve calculations to construct the model grid were performed by numerical one-dimensional radiation hydrodynamics code \texttt{STELLA} \citep{blinnikov1998,blinnikov2000,blinnikov2006}. \texttt{STELLA} evaluates the time evolution of the spectral energy distribution (SED), and we obtain a SED in each time step. The SED can be convolved with any transmission functions, and we can obtain theoretical light-curve models for any filters in addition to bolometric luminosity. The progenitor models with CSM are put into \texttt{STELLA} as an initial condition. The explosions are triggered as a thermal bomb by inserting thermal energy just above the mass cut set at 1.4~\Msun. The explosion energy ranges from 0.5~B ($1~\mathrm{B}\equiv 10^{51}~\mathrm{erg}$) to 5.0~B. Because \texttt{STELLA} does not treat explosive nucleosynthesis, the amount of \Ni\ is a free parameter. The \Ni\ mass is set from $0.001~\Msun$ to $0.3~\Msun$. \Ni\ is assumed to be mixed up to the half mass of the hydrogen-rich envelope of the progenitors.

\subsection{Observations}
We compare the observational data of SN~2023ixf presented in \citet{singh2024} with the model grid. The observational data presented in \citet{singh2024} ranges from the \textit{Swift} \textit{UVW2} band to the \textit{Kanata} $K_S$ band. In this work, we take the \textit{g} and \textit{r} bands for our analysis because SNe~II have been best studied in these bands in recent years thanks to ZTF \citep{bellm2019}. In addition, the CSM interaction strongly affects bluer bands, and the \textit{g} band light curve would contain the information on both the CSM interaction and recombination in the ejecta.

We also compare the pseudo-bolometric light curves from the observation and theory. The pseudo-bolometric light curve of SN~2023ixf is constructed using the flux information between 3250~\AA\ and 8900~\AA\ based on the observations reported in \citet{singh2024}. Since we have SED evolution in each timestep in our model grid, we can calculate the pseudo-bolometric luminosity in the same wavelength range for the model grid and compare it directly to the observed luminosity. 

We also use the information on the photospheric velocities from the model grid to compare with the observations. We compare the observed Fe~\textsc{ii} $\lambda 5169$ velocities measured by \citet{singh2024}.

\subsection{Comparison between the model grid and observations}

In this study, we evaluate the following $\chi^2$ value for all the light-curve models in the model grid, and we estimate the properties of SN~2023ixf based on the $\chi^2$ distribution. In our previous studies, we adopted a Bayesian approach to estimate the explosion properties of Type~II SNe (e.g., \cite{forster2018,subrayan2023,silva2024}). However, the previous studies used only a part of the model grid by, e.g., limiting the progenitor mass ranges because it has been challenging to incorporate all the models due to the demanding computational costs. Thus, we adopt a more straightforward approach to account for all the models in the grid.

The $\chi^2$ in this study is defined as
\begin{equation}
\chi^2 = \sum_i \frac{(x_{\mathrm{obs.},i}-x_{\mathrm{mod.},i})^2}{{\Delta x}^2},
\end{equation}
where $x_{\mathrm{obs.},i}$ is the $i$th observational data, $x_{\mathrm{mod.},i}$ is the model prediction for the observational data, and ${\Delta x}^2$ is an error. Although the errors in the observational data are small, we expect to have an unknown systematic error for the model predictions. Thus, we assign $\Delta x = 0.3~\mathrm{mag}$ for photometry and $\Delta x = 0.1~\mathrm{dex}$ for the pseudo-bolometric luminosity.

For each model, we evaluate $\chi^2$ by changing the explosion date of the model from 10~days to 0~day before the estimated explosion date of SN~2023ixf at $\mathrm{JD} = 2460083.315$ \citep{teja2023} with the interval of 0.1~days. The smallest $\chi^2$ value among the different explosion dates is then assigned to each model.

After obtaining $\chi^2$ for all the models, we rank the models based on the $\chi^2$ values. The model with the smallest $\chi^2$ is ranked as the number one, and the rank becomes higher as $\chi^2$ becomes higher. In this work, we estimate the properties of SN~2023ixf based on this rank distribution.

\section{Results}\label{sec:results}

Figure~\ref{fig:g} presents the cumulative rank distributions of the progenitor and explosion properties obtained using the \textit{g}~band light curves. The best fit model has the ZAMS mass of 10~\Msun, explosion energy of $2.0~\mathrm{B}$, $^{56}\mathrm{Ni}$ mass of 0.06~\Msun, mass-loss rate of $10^{-2.0}~\Msunpyr$, $\beta$ of 3.5, and CSM radius of $6\times 10^{14}~\mathrm{cm}$. The CSM mass of the model is 0.85~\Msun. The best-fit model is presented in Figure~\ref{fig:gbest}. Although we do not consider velocity information when we compare the model grid, we can find that the photospheric velocity evolution also matches well in the best-fit model.

When we look into the cumulative ranking distribution, the smaller ZAMS mass progenitors are preferred, and most of the best-fit models have the ZAMS mass of 10~\Msun.
Figure~\ref{fig:gzams} presents the best-fit models for each progenitor model. The higher mass models tend to have slow-declining light curves than those observed for SN~2023ixf.
Although the best-fit model has the explosion energy of $2.0~\mathrm{B}$, the models with $2.5~\mathrm{B}$ and $3.0~\mathrm{B}$ also tend to have small $\chi^2$ values. Thus, the explosion energy is estimated at $2.0-3.0~\mathrm{B}$. The most preferred $^{56}\mathrm{Ni}$ mass is 0.06~\Msun. The mass-loss rate of the best matching models are $10^{-2.0}~\Msunpyr$, but the models with $10^{-3.0}~\Msunpyr$ and $10^{-2.5}~\Msunpyr$ have relatively low ranks. Thus, the mass-loss rate can be estimated to be in the range of $10^{-2.0}-10^{-3.0}~\Msunpyr$. Although $\beta=\mathbf{3.5}$ gives the smallest $\chi^2$, the models with the other $\beta$ also provide a relatively low ranking, and we conclude that it is difficult to constrain $\beta$.
Since $\beta$ determines the CSM density structure at the immediate vicinity of the progenitor, the CSM mass is affected by $\beta$. If we fix the progenitor mass ($10~\Msun$), mass-loss rate ($10^{-2.0}~\Msunpyr$), and CSM radius ($6\times10^{14}~\mathrm{cm}$), the CSM mass ranges from $0.21~\Msun$ ($\beta = 0.5$) to $1.3~\Msun$ ($\beta = 5.0$). Thus, the CSM mass with the order of $0.1~\Msun$ is preferred, but the exact CSM density structure is not well constrained.
The models with the CSM radius of $6\times 10^{14}~\mathrm{cm}$ give the best-fit models, but the CSM radii of $8\times 10^{14}~\mathrm{cm}$ and $10^{15}~\mathrm{cm}$ provide the models with relatively low ranking.

So far, we have only looked into the comparison results obtained using the \textit{g}~band light curve. Figure~\ref{fig:gr} shows the cumulative ranking distribution obtained by using the \textit{g} and \textit{r}~band light curves. In this case, the best fit light-curve model has the ZAMS mass of 10~\Msun, explosion energy of $2.5~\mathrm{B}$, $^{56}\mathrm{Ni}$ mass of 0.06~\Msun, mass-loss rate of $10^{-3.0}~\Msunpyr$, $\beta$ of 5.0, and CSM radius of $10^{15}~\mathrm{cm}$. The CSM mass in this model is 0.52~\Msun. The best-fit light-curve model and its photospheric velocity evolution are presented in Figure~\ref{fig:grbest}. We find that adding the \textit{r}~band information does not change the results of the parameter estimations much. The best-fit parameters for the explosion energy, mass-loss rate, and $\beta$ change slightly, but the overall ranking distributions remain unchanged.

Finally, we show the comparison results based on the pseudo-bolometric light curve. The cumulative ranking distribution is shown in Figure~\ref{fig:bol}, and the best fit light-curve model is presented in Figure~\ref{fig:bolbest}. The best fit model has the ZAMS mass of 10~\Msun, explosion energy of $2.0~\mathrm{B}$, $^{56}\mathrm{Ni}$ mass of 0.04~\Msun, mass-loss rate of $10^{-3.0}~\Msunpyr$, $\beta$ of 4.5, and CSM radius of $10^{15}~\mathrm{cm}$. This model has a CSM mass of 0.44~\Msun. A large difference in the parameter estimations from those obtained by the \textit{g} and \textit{r}~band light curves is in the $^{56}\mathrm{Ni}$ mass. While most of the best-fit models based on the \textit{g} and \textit{r}~band light curves have the $^{56}\mathrm{Ni}$ mass of 0.06~\Msun, the best-fit models based on the pseudo-bolometric light curve has the $^{56}\mathrm{Ni}$ mass of 0.04~\Msun. The ranking distribution of the other parameters is not significantly different from those obtained by the \textit{g} and \textit{r}~band light curves.

\section{Discussion}\label{sec:discussion}
Based on the $\chi^2$ ranking distribution presented in the previous section, SN~2023ixf can be estimated to have the progenitor ZAMS mass of 10~\Msun, the explosion energy of $2.0-3.0~\mathrm{B}$, the $^{56}\mathrm{Ni}$ mass of $0.04-0.06~\Msun$, the mass-loss rate of $10^{-3.0}-10^{-2.0}~\Msunpyr$, and the dense CSM radius of $(6-10)\times 10^{14}~\mathrm{cm}$. The wind structure $\beta$ is not well constrained. These parameter estimations are obtained using the pre-computed model grid in \citet{moriya2023} without additional computations. Still, the estimated parameters are close to those obtained by detailed numerical investigations. For example, \citet{bersten2024} and \citet{martinez2023} estimated the ZAMS mass of 12~\Msun, the explosion energy of $1.2\times 10^{51}~\mathrm{erg}$, the $^{56}\mathrm{Ni}$ mass of 0.05~\Msun, the mass-loss rate of $3\times 10^{-3}~\Msunpyr$ with the wind velocity of $115~\kmps$, the dense CSM radius of $8\times 10^{14}~\mathrm{cm}$, and $\beta=5$ through the detailed numerical modeling. The estimated parameters are close to those we estimated in this paper. Our approach gives us a better idea of the uncertainties in the parameter estimation thanks to the $\chi^2$ ranking provided by the large model grid.

The parameter estimation approach we took in this paper is based on the pre-existing model grid. Thus, this paper does not aim to perfectly fit the light curves. As we can find in the best-fit models (Figs.~\ref{fig:gbest} and \ref{fig:grbest}), our light-curve models tend to decline slower than the observed light-curves, especially in the \textit{r}~band. The light-curve decline rate is closely related to the progenitor structure, including the hydrogen-rich envelope mass (e.g., \cite{dessart2013,moriya2016}). The assumed progenitor models from \citet{sukhbold2016} might not have been suitable for precisely reproducing the light-curve properties of SN~2023ixf. Still, the parameter estimation by the model grid can provide a good first guess for starting the detailed modeling of individual SNe.

Our models prefer a low-mass progenitor for SN~2023ixf. The progenitor mass of SN~2023ixf has been debated because the progenitor mass estimates from the pre-explosion images range from low mass (around $10~\Msun$, e.g., \cite{kilpatrick2023,pledger2023,vandyk2023,xiang2024}) to high mass (around $20~\Msun$, e.g., \cite{jencson2023,niu2023,qin2023,soraisam2023,liu2023}). Our mass estimate aligns with the low-mass progenitor estimates consistent with the results from the other light-curve modeling efforts (e.g., \cite{bersten2024}).
Our progenitor mass estimates are based on the progenitor model grid of \citet{sukhbold2016}. Adopting different progenitor models with different stellar evolution parameters can lead to different progenitor mass estimates due to inherent degeneracies with the parameters especially in the progenitor mass and radius (e.g., \cite{goldberg2019,goldberg2020}).

The properties of the dense, confined CSM estimated with our method are consistent with those obtained by other studies. The previous estimates based on the early-phase light-curve properties are around $10^{-3}-10^{-2}~\Msunpyr$ if we adopt the wind velocity of 10~\kmps\ as assumed in our model grid (e.g., \cite{teja2023,hiramatsu2023,jacobson-galan2023,zhang2023,zimmerman2023,soker2023}). The estimated mass-loss rate is also consistent with that estimated by comparing spectroscopic observations and models (e.g., \cite{jacobson-galan2023,bostroem2023}). 
The CSM density scales as $\rho_\mathrm{CSM}\propto \dot{M}/v_\mathrm{wind}$ (Eq.~\ref{eq:rho_csm}). In our model, we assume the terminal wind velocity of 10~\kmps. If we adopt the terminal wind velocity of 100~\kmps\ as observed in the high-resolution spectra of SN~2023ixf \citep{smith2023}, the mass-loss estimate increases by a factor of 10. However, the observed velocity of 115~\kmps might be affected by the radiative acceleration of the unshocked CSM (e.g., \cite{tsuna2023}).
The radius of the dense, confined CSM we estimated, $(6-10)\times 10^{14}~\mathrm{cm}$, is also consistent with other estimates (e.g., \cite{teja2023,jacobson-galan2023,hiramatsu2023,zhang2023,bostroem2023}), although the early-phase ultraviolet observations provide a smaller estimate of the CSM radius \citep{zimmerman2023}.

The mass-loss rate estimates based on the pre-explosion progenitor properties as well as X-ray observations are around $10^{-4}~\Msunpyr$ (e.g., \cite{grefenstette2023,chandra2023,jencson2023,niu2023,qin2023,soraisam2023}). This discrepancy likely originates from the aspherical nature of the dense, confined CSM identified by spectroscopic and polarimetric observations of SN~2023ixf (e.g., \cite{vasylyev2023,smith2023,singh2024}). Early light curves and spectra in the optical wavelengths are affected mainly by the dense part of the CSM.

\section{Conclusions}\label{sec:conclusions}
We have presented our results comparing the observed light curve of SN~2023ixf and the pre-existing synthetic light-curve model grid from \citet{moriya2023}. By simply evaluating the $\chi^2$ value for every model and rank the models based on the $\chi^2$ value, we estimated that SN~2023ixf have the progenitor ZAMS mass of 10~\Msun\ (adopting the progenitor models from \cite{sukhbold2016}), the explosion energy of $2.0-3.0~\mathrm{B}$, the $^{56}$Ni mass of $0.04-0.06~\Msun$, the mass-loss rate of $10^{-3.0}-10^{-2.0}~\Msunpyr$, and the dense, confined CSM radius of $(6-10)\times 10^{14}~\mathrm{cm}$. The wind structure parameter $\beta$ is not constrained well. The estimated parameters match those estimated by the previous detailed modeling of SN~2023ixf.

Our parameter estimation is based on the pre-existing model grid, and we did not perform any additional calculations to estimate the properties of SN~2023ixf. Thus, we can find some discrepancies between the best-matching models and observations, but our approach provides a reasonable parameter estimation. Our simple approach is sufficient to obtain a rough parameter estimation of Type~II SNe, building upon which we can perform more detailed modeling. In particular, such a simple, quick parameter estimation will be essential to characterize thousands of Type~II SNe discovered each year in the era of LSST. This study demonstrated that we can have a reasonable quick parameter estimation for them by using the pre-existing model grid and extracting information on the explosion and mass-loss mechanisms of RSGs.





\begin{ack}
The authors thank Jared Goldberg and the anonymous referee for constructive comments that improved this paper.
TJM is supported by the Grants-in-Aid for Scientific Research of the Japan Society for the Promotion of Science (JP24K00682, JP24H01824, JP21H04997, JP24H00002, JP24H00027, JP24K00668). This research was supported by the Australian Research Council (ARC) through the ARC's Discovery Projects funding scheme (project DP240101786). This work was supported by the JSPS Core-to-Core Program (grant number: JPJSCCA20210003).
\end{ack}




\bibliographystyle{apj}
\bibliography{pasj}

\end{document}